\documentclass{emulateapj}
\usepackage{natbib}
\usepackage[caption=false]{subfig}

\begin{document}
\title{The Gray Needle: Large Grains in the HD 15115 Debris Disk from LBT\footnotemark[1]/PISCES/\ks and LBTI\footnotemark[2]/LMIRcam/\lprime ~Adaptive Optics Imaging}
\footnotetext[1]{The LBT is an international collaboration among institutions in the United States, Italy and Germany. LBT Corporation partners are: The University of Arizona on behalf of the Arizona university system; Istituto Nazionale di Astrofisica, Italy; LBT Beteiligungsgesellschaft, Germany, representing the Max-Planck Society, the Astrophysical Institute Potsdam, and Heidelberg University; The Ohio State University, and The Research Corporation, on behalf of The University of Notre Dame, University of Minnesota and University of Virginia.}
\footnotetext[2]{The Large Binocular Telescope Interferometer (LBTI) is funded by the National Aeronautics and Space Administration as part of its Exoplanet Exploration program.}
\author{Timothy J. Rodigas\altaffilmark{1,2}, Philip M. Hinz\altaffilmark{1}, Jarron Leisenring\altaffilmark{3,4}, Vidhya Vaitheeswaran\altaffilmark{1}, Andrew J. Skemer\altaffilmark{1}, Michael Skrutskie\altaffilmark{3}, Kate Y.L. Su\altaffilmark{1}, Vanessa Bailey\altaffilmark{1}, Glenn Schneider\altaffilmark{1}, Laird Close\altaffilmark{1}, Filippo Mannucci\altaffilmark{5}, Simone Esposito\altaffilmark{5}, Carmelo Arcidiacono\altaffilmark{5}, Enrico Pinna\altaffilmark{5}, Javier Argomedo\altaffilmark{5}, Guido Agapito\altaffilmark{5}, Daniel Apai\altaffilmark{1}, Giuseppe Bono\altaffilmark{5}, Kostantina Boutsia\altaffilmark{6,7}, Runa Briguglio\altaffilmark{5}, Guido Brusa\altaffilmark{5}, Lorenzo Busoni\altaffilmark{5}, Giovanni Cresci\altaffilmark{5}, Thayne Currie\altaffilmark{8}, Silvano Desidera\altaffilmark{7}, Josh Eisner\altaffilmark{1}, Renato Falomo\altaffilmark{7}, Luca Fini\altaffilmark{5}, Kate Follette\altaffilmark{1}, Adriano Fontana\altaffilmark{6}, Peter Garnavich\altaffilmark{9}, Raffaele Gratton\altaffilmark{7}, Richard Green\altaffilmark{10}, Juan Carlos Guerra\altaffilmark{10}, J.M. Hill\altaffilmark{10}, William F. Hoffmann\altaffilmark{1}, Terry Jay Jones\altaffilmark{11}, Megan Krejny\altaffilmark{11}, Craig Kulesa\altaffilmark{1}, Jared Males\altaffilmark{1}, Elena Masciadri\altaffilmark{5}, Dino Mesa\altaffilmark{7}, Don McCarthy\altaffilmark{1}, Michael Meyer\altaffilmark{4}, Doug Miller\altaffilmark{10}, Matthew J. Nelson\altaffilmark{3}, Alfio Puglisi\altaffilmark{5}, Fernando Quiros-Pacheco\altaffilmark{5}, Armando Riccardi\altaffilmark{5}, Eleonora Sani\altaffilmark{5}, Paolo Stefanini\altaffilmark{5}, Vincenzo Testa\altaffilmark{5}, John Wilson\altaffilmark{3}, Charles E. Woodward\altaffilmark{11}, Marco Xompero\altaffilmark{5}}

\altaffiltext{1}{Steward Observatory, The University of Arizona, 933 N. Cherry Ave., Tucson, AZ 85721, USA}
\altaffiltext{2}{email: rodigas@as.arizona.edu}
\altaffiltext{3}{University of Virginia, Department of Astronomy, 530 McCormick Road, Charlottesville, VA  22903, USA}
\altaffiltext{4}{ETH Zürich, Institute for Astronomy, Wolfgang-Pauli-Strasse 27, 8093 Zürich, SWITZERLAND}
\altaffiltext{5}{INAF - Osservatorio Astrofisico di Arcetri, Largo E. Fermi 5, 50125 Firenze, ITALY}
\altaffiltext{6}{INAF - Osservatorio Astronomico di Roma, Via Frascati 33 00040 Monte Porzio Catone, ITALY}
\altaffiltext{7}{INAF – Osservatorio Astronomico di Padova, Vicolo dell’ Osservatorio 5, I-35122, Padova, ITALY}
\altaffiltext{8}{NASA-Goddard Space Flight Center, Greenbelt, MD, USA}
\altaffiltext{9}{Department of Physics, 225 Nieuwland Science Hall, University of Notre Dame, Notre Dame, IN 46556, USA}
\altaffiltext{10}{LBT Observatory, University of Arizona, Tucson, AZ 85721, USA}
\altaffiltext{11}{Institute for Astrophysics, University of Minnesota, 116 Church St. SE, Minneapolis, MN 55455, USA}

\newcommand{\about}{$\sim$~}
\newcommand{\mj}{M$_{J}$}
\newcommand{\degrees}{$^{\circ}$}
\newcommand{\arcseconds}{$^{\prime \prime}$}
\newcommand{\asec}{$\arcsec$}
\newcommand{\fasec}{$\farcs$}
\newcommand{\lprime}{$L^{\prime}$}
\newcommand{\ks}{$Ks$~}
\newcommand{\mjyasec}{mJy/arcsecond$^{2}$}
\newcommand{\microns}{$\mu$m}

\shorttitle{Large Grains in the HD 15115 Debris Disk}
\shortauthors{Rodigas et al.}

\begin{abstract}
We present diffraction-limited \ks band and \lprime ~adaptive optics images of the edge-on debris disk around the nearby F2 star HD 15115, obtained with a single 8.4 m primary mirror at the Large Binocular Telescope. At \ks band the disk is detected at signal-to-noise per resolution element (SNRE) \about 3-8 from \about 1-2\fasec 5 (45-113 AU) on the western side, and from \about 1.2-2\fasec 1 ~(63-90 AU) on the east. At \lprime ~the disk is detected at SNRE \about 2.5 from \about 1-1\fasec 45 (45-90 AU) on both sides, implying more symmetric disk structure at 3.8 \microns . At both wavelengths the disk has a bow-like shape and is offset from the star to the north by a few AU. A surface brightness asymmetry exists between the two sides of the disk at \ks band, but not at \lprime . The surface brightness at \ks band declines inside 1\asec ~(\about 45 AU), which may be indicative of a gap in the disk near 1\asec . The \ks - \lprime ~disk color, after removal of the stellar color, is mostly grey for both sides of the disk. This suggests that scattered light is coming from large dust grains, with 3-10 \microns -sized grains on the east side and 1-10 \microns ~dust grains on the west. This may suggest that the west side is composed of smaller dust grains than the east side, which would support the interpretation that the disk is being dynamically affected by interactions with the local interstellar medium. 
\end{abstract}

\section{Introduction}
Debris disks are thought to be signposts for planet formation since their reservoirs of dust grains must be frequently replenished by collisions between larger, planetesimal-sized bodies \citep{wyatt08}. The idea of debris disks as markers for planets has been supported by the recent direct detections of wide-orbiting, massive exoplanets in systems that also harbor luminous debris disks (e.g., \cite{hr87994thplanet,kalas,betapic}). In systems with spatially resolved disks where no planets are detected, knowledge of dust grain sizes, compositions, density distributions, and disk morphologies can help us infer where any unseen planets might reside \citep{diskgapplanets,quillenfomalhaut}. 

To date the highest resolution images of debris disk scattered light have been obtained in the visible to near-infrared (NIR) with the Hubble Space Telescope (HST) and from the ground with adaptive optics (AO). Unfortunately, due to thermal emission from the Earth's atmosphere, obtaining ground-based detections of scattered light from extended debris disks at wavelengths longward of 2 \microns ~is difficult. But the 2-5 \microns ~wavelength region is a window into important debris disk properties. Scattered light at these wavelengths is sensitive to larger grains ($\gtrsim$ a few \microns ) than can be probed by visible to NIR scattered light. Imaging at these wavelengths can also constrain dust grain compositions, in particular the fraction of water ice on the surface of dust grains \citep{inoue}. The 3-4 \microns ~wavelength region is also particularly sensitive to the thermal radiation of massive exoplanets (e.g., \cite{burrows,hinz2,hr87994thplanet,betapic}). Therefore imaging at these wavelengths simultaneously probes disk scattered light from large grains and thermal radiation from exoplanets. 

Because debris disks are optically thin, edge-on disks have a larger optical depth along the line of sight than face-on disks. Therefore, in the favorable case of an edge-on inclination, disk flux densities (assuming isotropic scattering) are maximized, facilitating their detection above the high sky background. In addition, when creating a reference point spread function (PSF) from the intrinsic rotation of the sky relative to the telescope using the science target (as in \cite{liuaumic} and \cite{aumic}, for example), edge-on disks are more resistant to disk self-subtraction when differencing the reference PSF from the observed target. However being edge-on and bright does not guarantee detections longward of 2 \microns , as \cite{aumic} obtained confident detections of the AU Mic debris disk in the $J$, $H$, and \ks bands, but reached only a 3$\sigma$ upper limit of \about 12.6~mags/arcsecond$^2$ at \lprime ~(in 12 minutes of integration). To confidently detect scattered light at 3.8 \microns ~from the ground, longer integrations (relative to integrations at shorter wavelengths) are required to mitigate the higher sky background. Additionally, detections are facilitated if the disk consists of large, grey-scattering grains as opposed to small, blue-scattering grains.

HD 15115 is a nearby (d = 45.2 $\pm$ 1.3 pc \citep{van}) F2 star with an edge-on asymmetric debris disk previously spatially resolved in the visible and NIR \citep{kfg,debes}. The star is believed to be young for several reasons: it has shared kinematics with  the 12 Myr old $\beta$ Pic moving group \citep{moordisks06}, it is believed to be on the zero-age main sequence (Eric Mamajek, private communication), and it has a high fractional luminosity circumstellar disk ($f_{d}$ = 4.9 $\times$ 10$^{-4}$), which is more commonly seen for younger stars \citep{moordisks06}. However other indicators, such as Ca II H and K lines and X-ray emission, may point to a much older age, perhaps 100-500 Myr \citep{silverstone,rhee}. Furthermore \cite{debes} refuted the evidence for the star being a $\beta$ Pic group member, based on backtracking the star's proper motion and radial velocity. Given the large uncertainty in this star's age, for the purposes of this paper we take the star's age to be conservatively between 10 Myr and 1 Gyr.

HD 15115's circumstellar disk is believed to be gas-poor, with $<$ 0.28x10$^{-4}$ M$_{\oplus}$ in CO gas \citep{moorgas}, and has 0.047 M$_{\oplus}$ in dust mass \citep{zuck04,williams06}. Therefore the disk is considered to be predominantly ``debris." At visible wavelengths, the disk is highly asymmetric, with the western lobe extending out to a stellocentric radius of \about 12\asec ~in a ``needle"-like feature \citep{kfg}. \cite{kfg} reported blue F606W - $H$ colors, especially at large separations (hence the ``blue needle"). \cite{debes} also saw blue scattering beyond 2\asec , but reported red F110W - $H$ colors at 1\asec . The red color of the disk close to the star makes HD 15115 an attractive target for 2-4 \microns ~imaging.

Resolving the disk longward of 2 \microns ~would allow us to constrain the population of large grains in the disk, especially closer to the star. This requires high-Strehl ratio, low thermal background observations. From the ground, we need a precise adaptive optics (AO) system and a minimal number of warm surfaces in the optical path. 

The Large Binocular Telescope (LBT) satisfies these requirements. Currently it has a single adaptive secondary mirror with 675 actuators, capable of operating with up to 500 modes of correction on one of the two 8.4 m primary mirrors. This allows for high-Strehl ratio (up to 70-80$\%$ at $H$ band and up to 95$\%$ at \ks band), low thermal background observations (see \cite{LBTAO} and references therein for a review of the LBT AO system). 

We observed HD 15115 at \ks band and at \lprime ~with the DX (right) primary mirror and its adaptive secondary mirror at the LBT. In Section 2 we describe the observations we carried out at \ks band with PISCES \citep{pisces} and at \lprime ~with LMIRcam \citep{lmircam} combined with the non-interferometric mode of the Large Binocular Telescope Interferometer (LBTI, \cite{lbti}), as well as our data reduction methods. In Section 3 we present our results on the disk structure and surface brightness (SB) profiles, and limits on planets in the system. In Section 4 we discuss the implications of our results, including the disk color and grain sizes. In Section 5 we summarize the main results. 

\section{Observations and Data Reduction}
\subsection{LBT/PISCES \ks band}
We carried out our \ks band observations of HD 15115 on UT November 9 2011 with the PISCES camera, a high-contrast 1-2.5 \microns ~imager modified for use at the LBT. PISCES has a field of view (FOV) of \about 19\fasec 5 on a side and a plate scale of 19.4 mas/pixel. PISCES, which uses a pyramid wavefront sensor with natural starlight, was mounted at the right front bent Gregorian focus behind the First Light Adaptive Optics (FLAO) system \citep{flao}. Skies were clear and the seeing was good (0.5-1\asec ) for most of the duration of the observations. We used the single 8.4 m DX (right) mirror combined with its adaptive secondary mirror. The camera rotator was fixed at a fiducial position to allow for angular differential imaging (ADI; \cite{adi}), and no coronagraphs were used. We obtained 603 images with 4 s integration per image, resulting in a total integration of 40.2 minutes and \about 40\degrees ~of parallactic angle rotation. For approximately the first half of the integration, the star remained stationary on the detector and was not nodded or dithered. For the second half of the data, the star was moved to the opposite side of the detector, after approximately half an hour of down time during which the AO was not functioning. We also obtained several 0.8 s unsaturated images of HBC 388 in the narrowband BrG ($\lambda_{0}$ = 2.169 \microns ) filter, a few hours after HD 15115, as a photometric reference.

The raw images were corrected for cross-talk and geometrical distortion\footnotemark[1] \footnotetext[1]{see http://aries.as.arizona.edu/ and links therein}. After these steps, all data reduction was performed with custom Matlab scripts. Each half of the data was used to subtract the sky and detector artifacts from the opposite half, and flat-field corrections were applied from sky images obtained earlier in the night. Each image was divided by its exposure time to obtain units of counts/s. All images were saturated at the cores (inside $\lambda$/D) of the PSF, so the star positions were determined by calculating the center of light using an annulus from just outside the saturated region to 2\asec . We performed similar center of light calculations on the unsaturated images of HBC 388 and found the difference between the true centroid and the calculated centroid to be \about 1 pixel (0\fasec 194). The images were then registered to a fiducial position. For each image, we calculated the azimuthal median radial profile after masking out the spider arms and other detector artifacts, constructed an image built from the radial profile, and subtracted that from the original image. 

After median-combining the first half of the images, the second half of the images, and all images together, we determined that the second half of the data was of much poorer quality than the first half, perhaps due to the worse seeing during this part of the night. Consequently we only used the first half of the data in our final data reduction and analysis. We constructed the master PSF reference by median-combining the first 284 images, and scaling and subtracting this reference image from each of the first 284 images. The scale factor was determined by minimizing the flux after subtraction inside a 0.15-1\asec ~annulus (outside the saturated region). Each image was then flipped about the vertical axis (due to a mirror flip inside PISCES) and rotated by its parallactic angle plus an additional rotational offset to obtain North-up, East-left. The offset was determined to be 80.88\degrees ~(the position angle of the detector) plus 3.9\degrees ~(see Section 3.2 for a discussion of how the 3.9\degrees ~offset was determined). The total parallactic angle rotation for this set of images was approximately 20\degrees .

\begin{figure}[h]
\centering
\subfloat[]{\label{fig:ksimage}\includegraphics[scale=0.43]{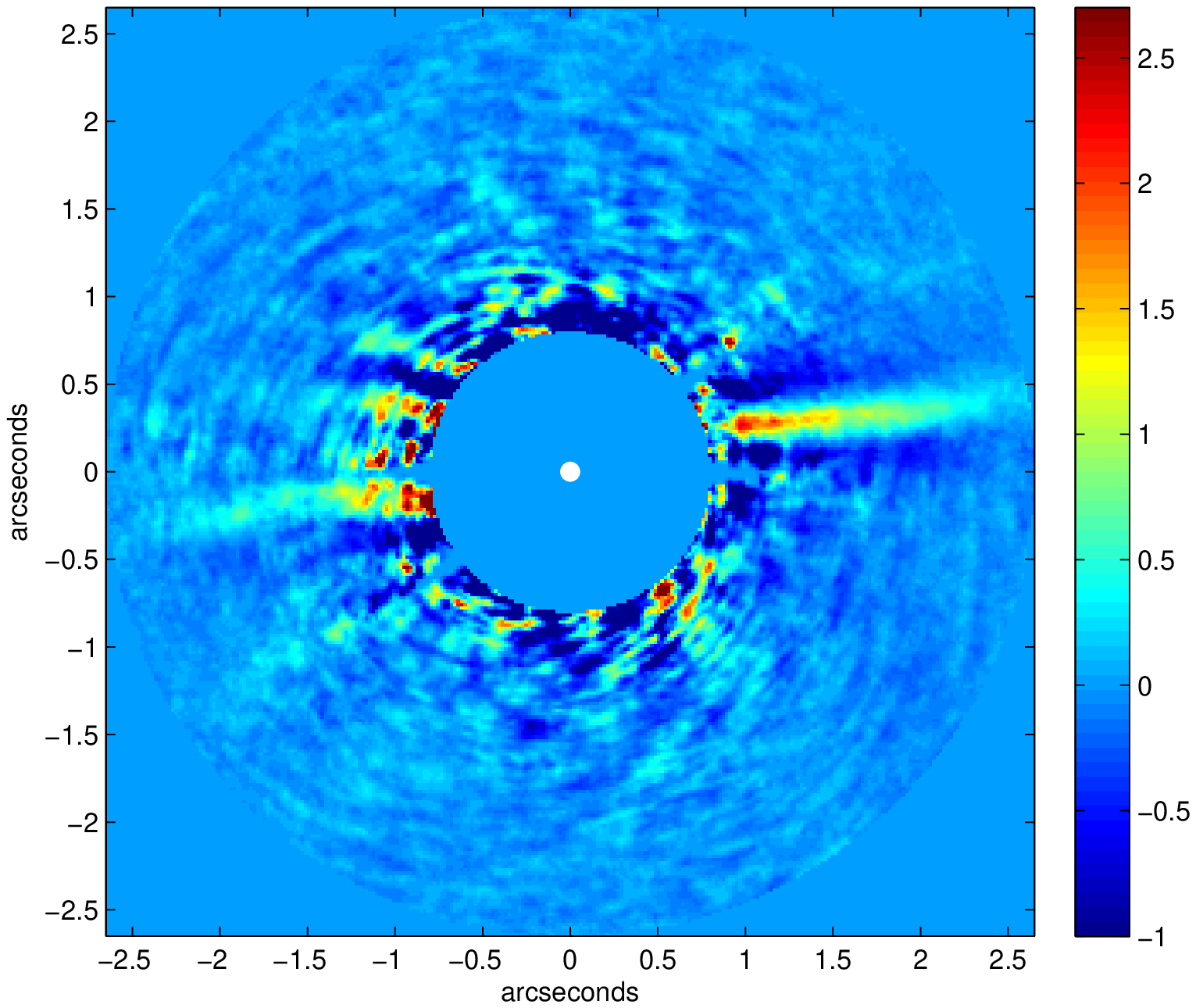}} \\
\subfloat[]{\label{fig:ksnre}\includegraphics[scale=0.43]{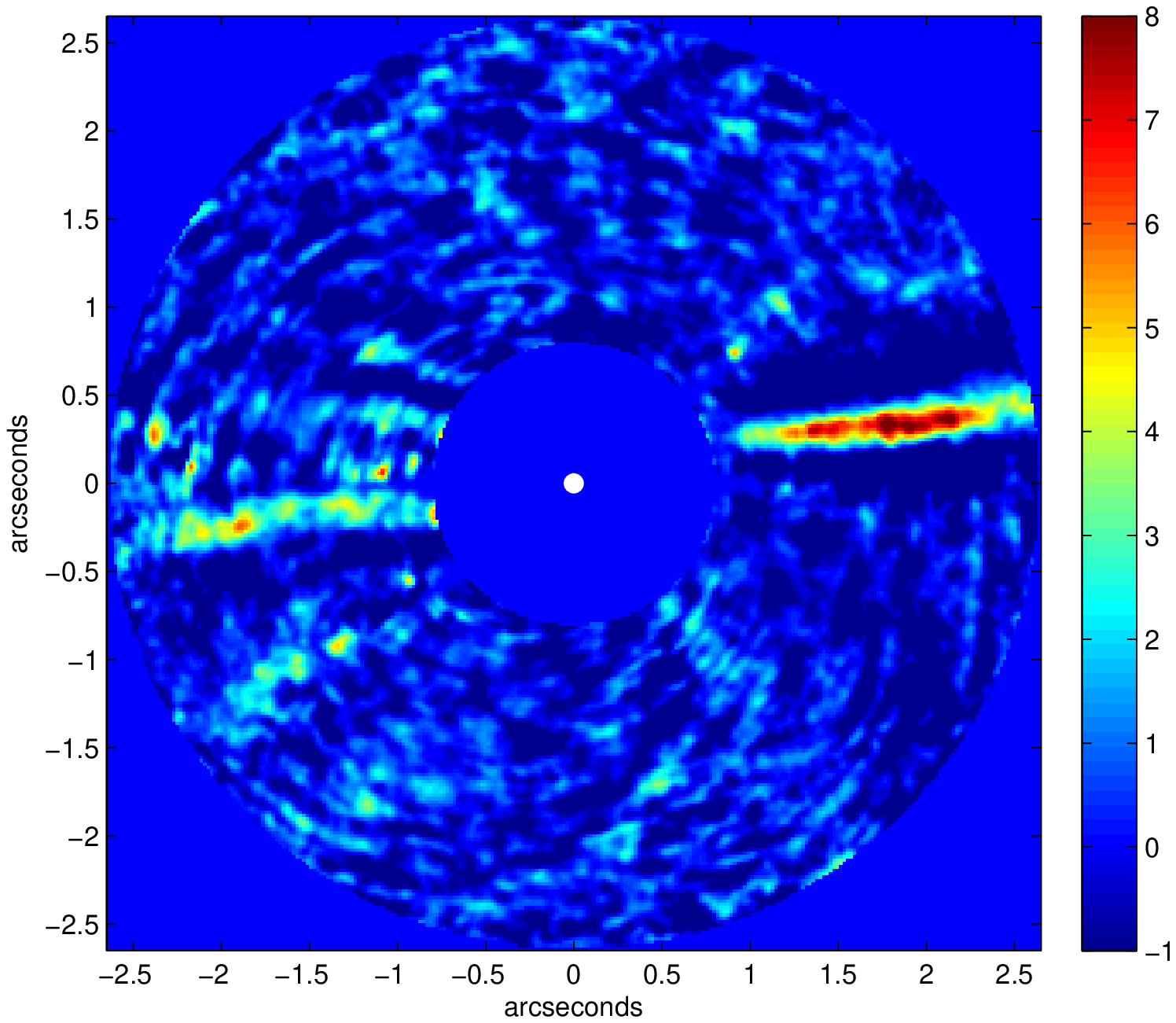}}
\caption{\textit{Top}: Final PISCES \ks band image of the HD 15115 debris disk, in units of \mjyasec , with North-up, East-left. The white dot marks the location of the star, and represents the size of a resolution element at \ks band. For display purposes a 0\fasec 8 radius mask has been added in post-processing. The western side SB is \about a magnitude/arcsecond$^{2}$ brighter than the eastern side SB, as is seen at shorter wavelengths \citep{kfg,debes}. \textit{Bottom}: SNRE map of the image. The eastern side is detected at SNRE \about 3 out to \about 2\fasec 1. The western side of the disk is detected at SNRE \about 3-8 from \about 1-2\fasec 5.}
\end{figure}

We median-combined the PSF-subtracted, rotated images, masking out the residual streaks left by the spider arms and other detector artifacts. The final image revealed HD 15115's edge-on asymmetric disk structure at the expected position angle (PA) of \about 279\degrees ~\citep{kfg,debes}). To obtain the highest SNR image, we reduced the first 284 (better quality) images using a conservative LOCI \citep{loci} algorithm, as in \cite{moth} and \cite{thalmannhr4796}. Specifically, to preserve disk flux we required a field rotation of at least 3 times the full-width half-max (FWHM) at \ks band (= 3 $\times$ 0\fasec 0525) between images. We also set the optimization section ($N_{A}$) to 2000 to ensure less self-subtraction. Fig. \ref{fig:ksimage} shows our final PISCES/\ks band image in \mjyasec. 

Because we could not obtain any unsaturated images of HD 15115 (the shortest integration time possible with PISCES is 0.8 s, which saturates the star), the flux calibration for obtaining units of \mjyasec ~was calculated as follows: we scaled the ratio of the PSF halo of HBC 388 to its total flux and compared that to the halo of the median-combined PSF reference image of HD 15115 (so that the disk flux contribution was washed out). The ratio was determined for different halo sizes, ranging from 0\fasec 3 to 1\asec , 1\fasec 25, and 1\fasec 5--all yielding the same ratio to within less than 1$\%$. We adopted this method, as opposed to scaling by the relative bandwidths in the \ks ~and BrG filters, because the latter method may be less accurate due to changing photometric conditions.

Fig. \ref{fig:ksnre} shows the map of the signal-to-noise per resolution element (SNRE) of the final \ks band image. The SNRE was determined by masking out the disk in the final image, smoothing the image by a Gaussian kernel with FWHM = 0\fasec 0525 (= $\lambda$/D), calculating the standard deviation of the smoothed image in concentric annuli around the star, then dividing this noise image into the final smoothed \ks band image. From the SNRE map, the disk is detected on the eastern side at SNRE \about 3 between 1.2-2\fasec 1; and on the western side the disk is detected at SNRE \about 3-8 between 1-2\fasec 5.

We refer the reader to the Appendix, wherein we describe how we inserted an artificial model disk into the raw data and re-reduced the data using the LOCI algorithm, to understand LOCI's effects on disk position angle, FWHM, and surface brightness as a function of distance from the star. The algorithm's effects on these values were measured and accounted for in the disk analysis found later in this paper.

\begin{figure}[h]
\subfloat[]{\label{fig:Limage}\includegraphics[scale=0.43]{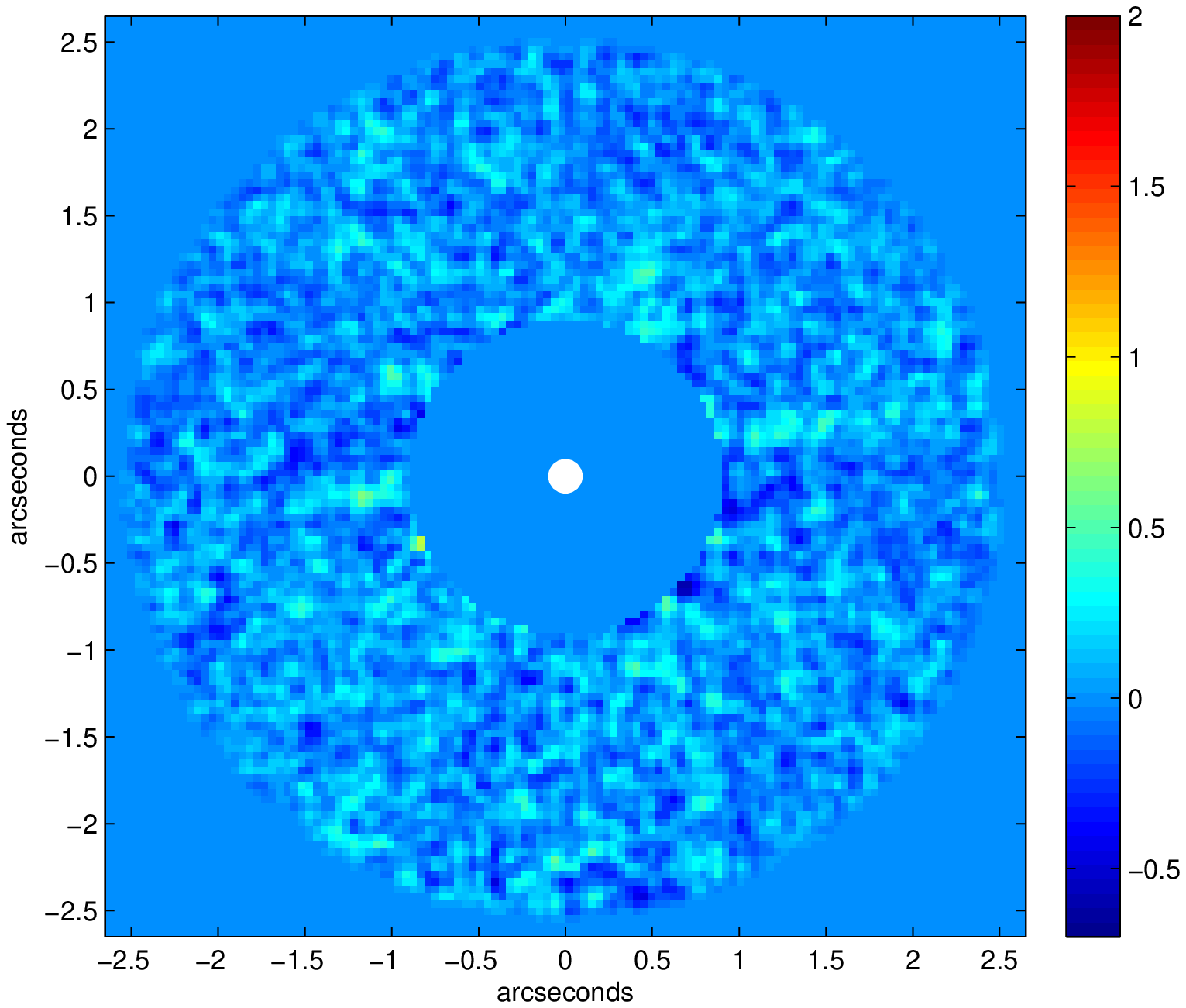}} \\
\subfloat[]{\label{fig:LSNR}\includegraphics[scale=0.43]{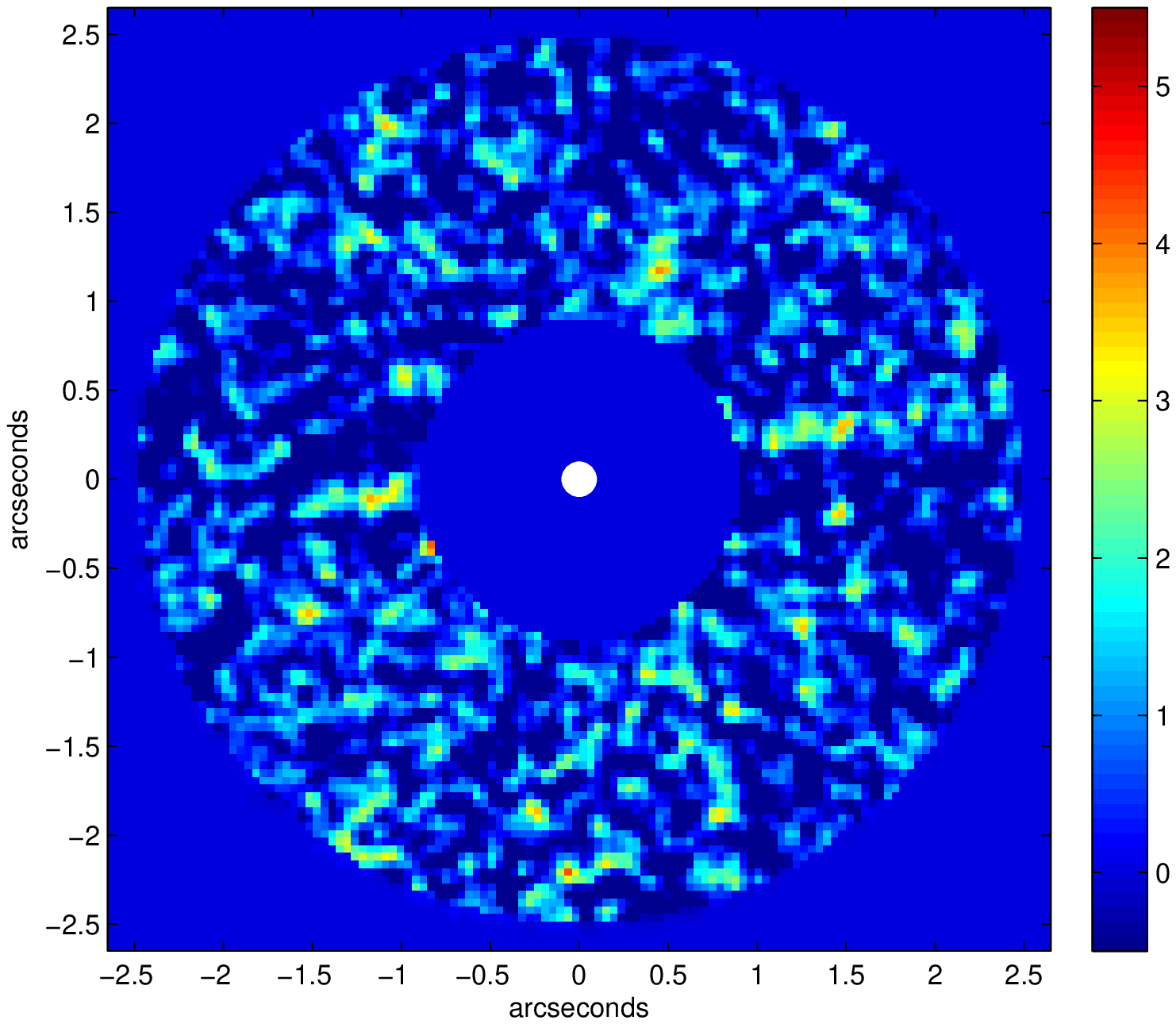}} \\
\caption{\textit{Top}: Final \lprime ~image of the disk obtained with LMIRcam, in units of \mjyasec , with North-up, East-left. The white dot marks the location of the star, and represents the size of a resolution element at \lprime . For display purposes a 0\fasec 9 radius mask has been added in post-processing. The image has been smoothed by a Gaussian kernel with FWHM $= \lambda$/D and binned by a factor of 4 to bring out the disk. At \lprime ~the disk is mostly symmetric and is equally bright on both sides. These features are nearly opposite to what is seen at \ks band and at shorter wavelengths. \textit{Bottom}: SNRE map of the final \lprime ~image shown above. The disk is detected at SNRE \about 2.5 between \about 1-1\fasec 5 on both sides.}
\end{figure}

\subsection{LBTI/LMIRcam \lprime}
We carried out our \lprime ~observations of HD 15115 on UT November 16 2011 with LMIRcam \citep{lmircam}, a high-contrast 3-5 \microns ~imager designed for use with LBTI \citep{lbti}. These observations were made in direct imaging mode, without any interferometric combination. Like the FLAO system, LBTI also uses a pyramid wavefront sensor with natural starlight for the adaptive reference source. We used the 8.4 m DX mirror combined with its adaptive secondary. LMIRcam has a FOV of \about 11\asec ~on a side and a plate scale of 10.7 mas/pixel, determined from observations of the binary star HD 37013 the previous night. LBTI is fixed to the telescope and thus cannot rotate, facilitating the use of ADI throughout the observations. We obtained 2319 images, the first 5 with integrations of \about 10 seconds, and the remainder with integrations of \about 2.5 seconds. The total integration obtained on the source was therefore \about 97 minutes. Observing conditions were good, with clear skies and \about 1\asec ~seeing. For much of the second half of the integration, the wind picked up and caused some vibration-induced blurring of the saved images. At the end of the acquisition, we obtained several short-exposure, unsaturated images of HD 15115 and HD 40335, an \lprime ~standard star \citep{leggett}, for photometric references. Throughout the long exposure images, each image was saturated inside $\lambda$/D. Due to a poorly-mounted dichroic, astigmatism in the PSF was visible inside the central \about 0\fasec 2 of each image, creating a ``cross" pattern on the PSF. This did not ultimately have any measurable effect on the quality of the data taken, and the instrumental astigmatism was corrected after the observing run. Throughout all observations, we nodded the star diagonally by several arcseconds, and dithered the star by 0\fasec 25 in a box around the center of each nod position.

All data reduction was performed in custom Matlab scripts. Images were divided by their individual integration times to obtain units of counts/s, nod-subtracted, and corrected for bad pixels. Due to the large quantity of data, images were averaged in sets of 4. We registered the images to a fiducial pixel by calculating the center of light around the star, excluding the saturated pixels within \about $\lambda$/D, in the same manner as for the \ks band images. We performed similar center of light calculations on the unsaturated images of HD 15115 and found the difference between the measured and true centroid to be \about 0.2 pixels (2 mas). We then calculated the radial profile of each image and subtracted it, also in the same manner as for the \ks band images. To test the quality of the data, we median-combined the first half of the data, the second half, and the entire set. We saw that though the quality of the PSFs was good, the second half of the data suffered from bright extended streaks appearing in the raw data, out to several arcseconds. These streaks were not caused by the spider arms, and were not symmetric, so they could not be easily masked out and would overwhelm the brightness of any disk structure seen in the final image. Therefore we used only the first half of the data (\about 40 minutes of integration, with \about 50\degrees ~of parallactic angle rotation). 

After scaling and subtracting (in the same manner as for the \ks band images) the master PSF from each image in the first half of the data (240 averaged images; 960 total), we rotated each image by its parallactic angle plus an offset to obtain North-up, East-left. The offset was determined to be 1.81\degrees ~$\pm$ 0.0685\degrees ~by calculating the PA of the binary HD 37013, which was observed with LMIRcam the night before. The error was determined by independently calculating the offset with two different software routines and measuring the difference. The PSF-subtracted images were then median-combined. The final image, after smoothing by a Gaussian kernel with FWHM = $\lambda$/D and binning by a factor of 4, revealed edge-on disk structure at a similar PA to the disk in the PISCES \ks band image. 

To obtain the highest-SNR image, we reduced the data using a conservative LOCI algorithm. Here we required 3 FWHMs (= 3 $\times$ 0\fasec 1) of rotation between images, and the optimization section $N_{A}$ was 300 (\about 1\fasec 5). Fig. \ref{fig:Limage} shows this final image in \mjyasec , where the conversion to \mjyasec ~was determined using the unsaturated images of HD 15115. Fig. \ref{fig:LSNR} shows the SNRE map of the final \lprime ~image, where the SNRE map was calculated in the same manner as for the \ks band image. The disk at \lprime ~is detected at SNRE \about 2.5 from 1-1\fasec 5 on both sides of the disk.

We refer the reader to the Appendix, wherein we describe how we inserted an artificial model disk into the raw data and re-reduced the data using the LOCI algorithm, to understand LOCI's effects on disk position angle, FWHM, and surface brightness as a function of distance from the star. The algorithm's effects on these values were measured and accounted for in the disk analysis found later in this paper.

\subsection{Ancillary Keck/NIRC2 \ks band data}
During the LBT observations, the PISCES camera lacked an accurate astrometric solution to determine the PA of true North on the detector. This is largely due to faulty fits header reporting of the detector PA, as well as the lack of observations of a standard binary for astrometric calibration. Therefore we compared our PISCES final image to \ks band data of HD 15115 obtained on the night of UT August 12 2011 with Keck/NIRC2, which has a very precise astrometric calibration and distortion solution determined from repeated observations of the galactic center \citep{yelda}.

Briefly, the data were taken with the narrow camera with the 0\fasec 6 diameter coronagraphic mask and consisted of coadded 10 s exposures, with a cumulative integration time of \about 1280 seconds and total field rotation of \about 39\degrees . The data were processed using a conservative LOCI algorithm. To provide a detection of the HD 15115 disk minimally-biased by LOCI processing, we adopted relaxed LOCI parameters comparable to those used in \cite{thalmannhr4796}. Specifically, we set the LOCI parameter for the required rotation gap, $\delta$, to be 3 times the FWHM (3 $\times$ 0\fasec 052) and set the optimization area, $N_{A}$, to be 3000 (see \cite{curriehr8799} for additional details on the algorithm used). The disk was detected between \about 1-2\asec ~at SNRE \about 3-5. The final image was rotated clockwise by an additional 0.252\degrees ~to obtain North-up, East-left. We show only the PISCES \ks band image in this paper because the disk in the PISCES image was detected at higher SNRE. The Keck/NIRC2 data will be discussed and analyzed in Currie et al. (2012, in prep.). 

\section{Results}
\subsection{Disk FWHM}
To better define the \ks band morphology of the disk, we measured the disk FWHM perpendicular to the major axis of the disk as a function of stellocentric distance. We did this as follows: first we rotated the final image by 7\degrees ~so that the disk midplane was horizontal; next at each discrete horizontal (pixel) distance from the star we located the brightest pixel in the column of pixels perpendicular to the disk; we then placed a 7 pixel by 25 pixel (0\fasec 13 by 0\fasec 48) box centered on the brightest pixel, and summed up the counts/pixel along each row of the box, producing a 1D array of 25 values. Next we fit a Gaussian to this array. The Gaussian fit gives the disk midplane pixel location along with the FWHM at that location. We also measured the disk FWHM as the number of pixels along each row with SNRE $>$ 3. We took the final disk FWHM value as the average of the Gaussian-fitted value and the SNRE-width value, with the error being half the difference between the two. We also added in the constant FWHM offset (\about 0\fasec 03) determined from insertion and recovery of an artificial disk in the raw data (see Appendix for details). 

Fig. \ref{fig:EFWHM} and Fig. \ref{fig:WFWHM} show the east and west FWHM of the disk as a function of distance from the star, respectively. The western disk FWHM increases gradually with increasing distance from the star, though this may be an effect from the LOCI reduction (see Appendix), while no strong correlation is evident for the eastern FWHM. The median FWHM across the entire disk was found to be 0\fasec 21 $\pm$ 0\fasec 03. The intrinsic FWHM of the disk is computed by subtracting off in quadrature the PSF broadening at \ks band (= 0\fasec 0525), but this has a negligible effect on the disk FWHM, yielding the same value. Within the uncertainty, this value is consistent with both the intrinsic disk FWHM at 1.1 \microns , 0\fasec 26 $\pm$ 0\fasec 07, where we computed this value by averaging the reported intrinsic disk FWHM of each side of the disk \citep{debes}, as well as the FWHM of the disk at 0.6 \microns , reported by \cite{kfg} as 0\fasec 19 $\pm$ 0\fasec 1. The measured disk FWHM at \ks band being $>$ 3 times the size of the PSF FWHM (0\fasec 0525) suggests that we have spatially resolved the disk. 
\begin{figure}[t]
\centering
\subfloat[]{\label{fig:EFWHM}\includegraphics[scale=0.45]{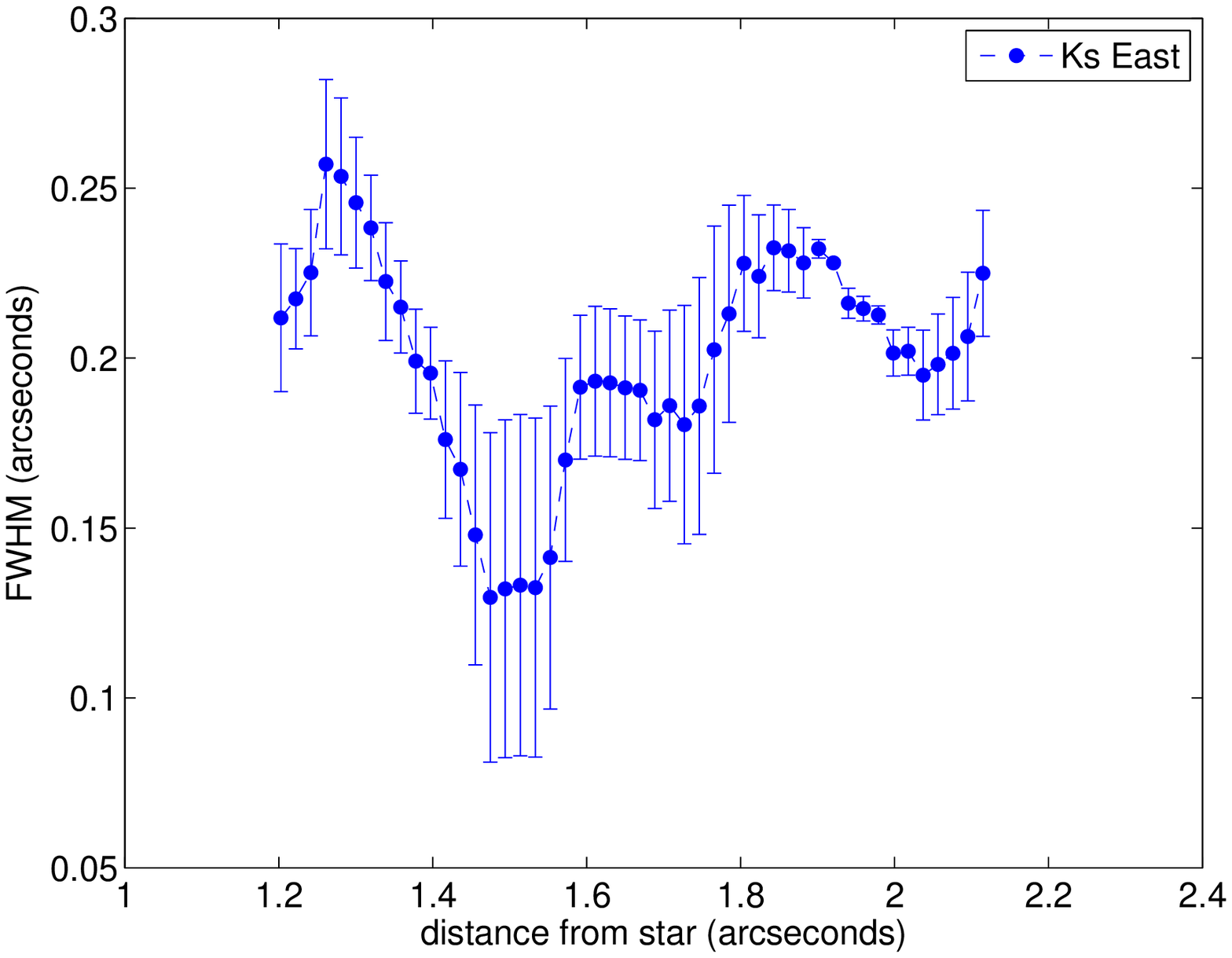}} \\
\subfloat[]{\label{fig:WFWHM}\includegraphics[scale=0.45]{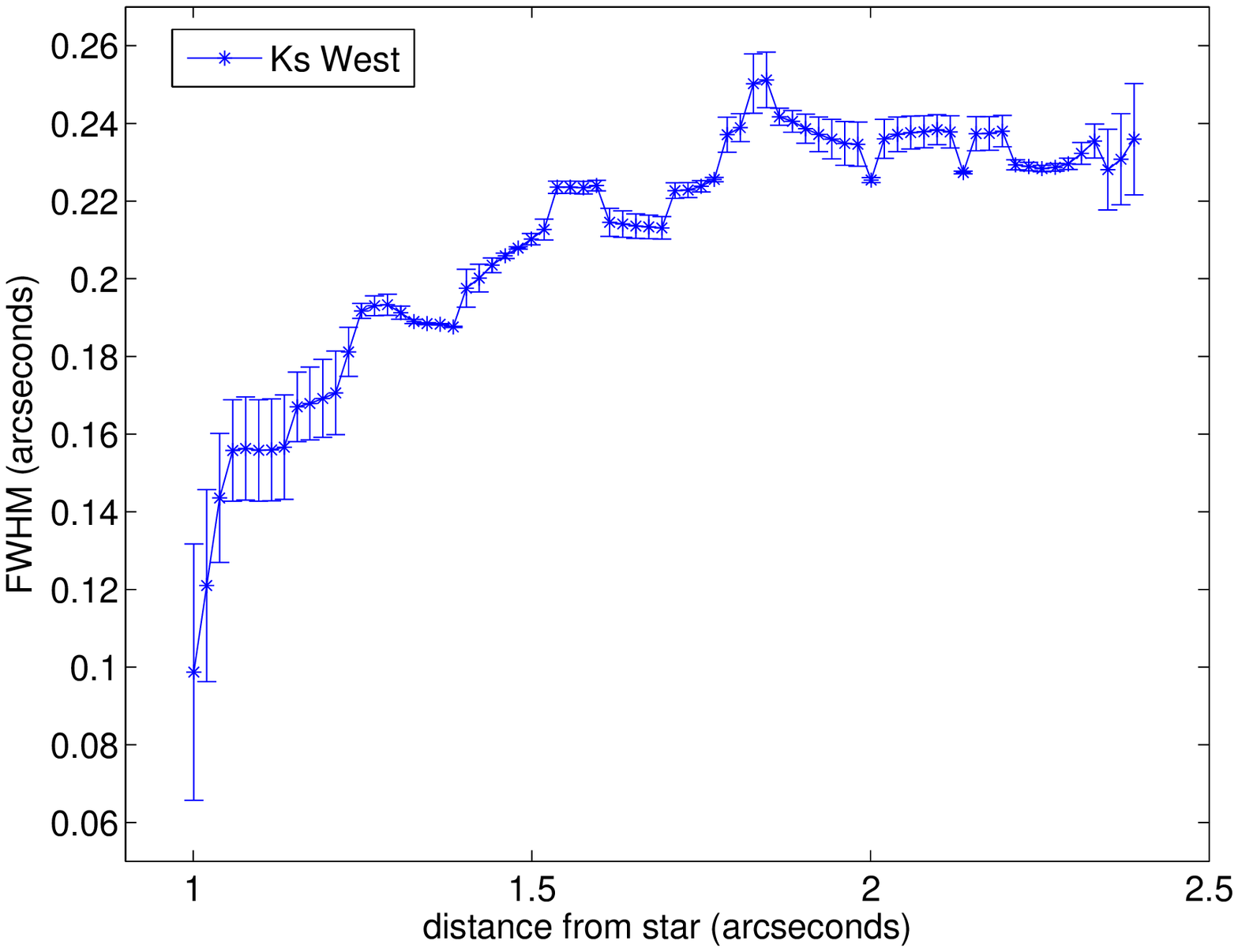}} \\
\caption{\textit{Top}: \ks band eastern disk FWHM as a function of distance from the star. See text for methodology. There is no obvious trend in FWHM vs. stellocentric distance. \textit{Bottom}: Same as above, except for the western side of the disk. The FWHM increases with increasing distance from the star, though this may be an effect from the data reduction (see Appendix).}
\end{figure}
\begin{figure}[t]
\centering
\subfloat[]{\label{fig:EPA}\includegraphics[scale=0.45]{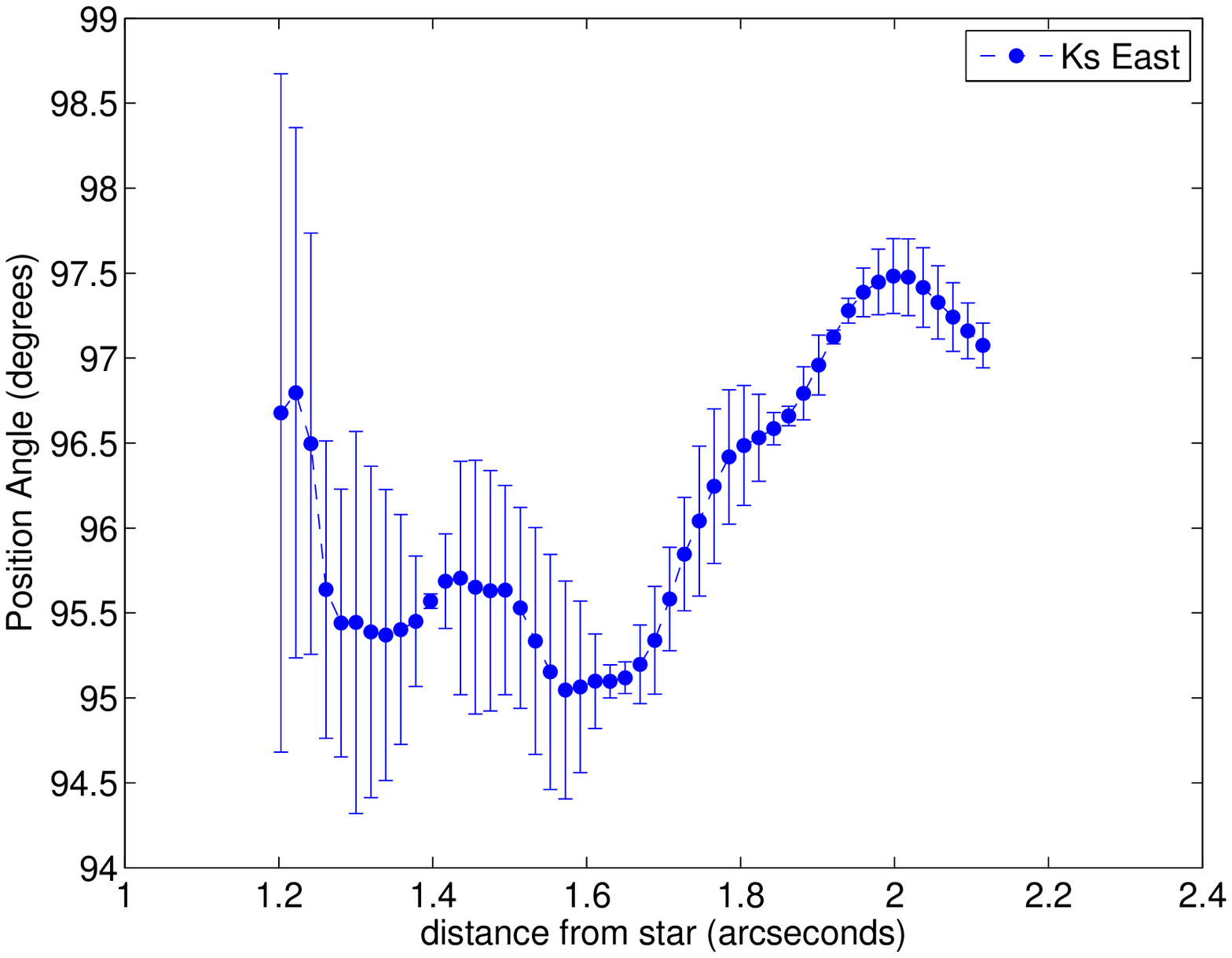}} \\
\subfloat[]{\label{fig:WPA}\includegraphics[scale=0.45]{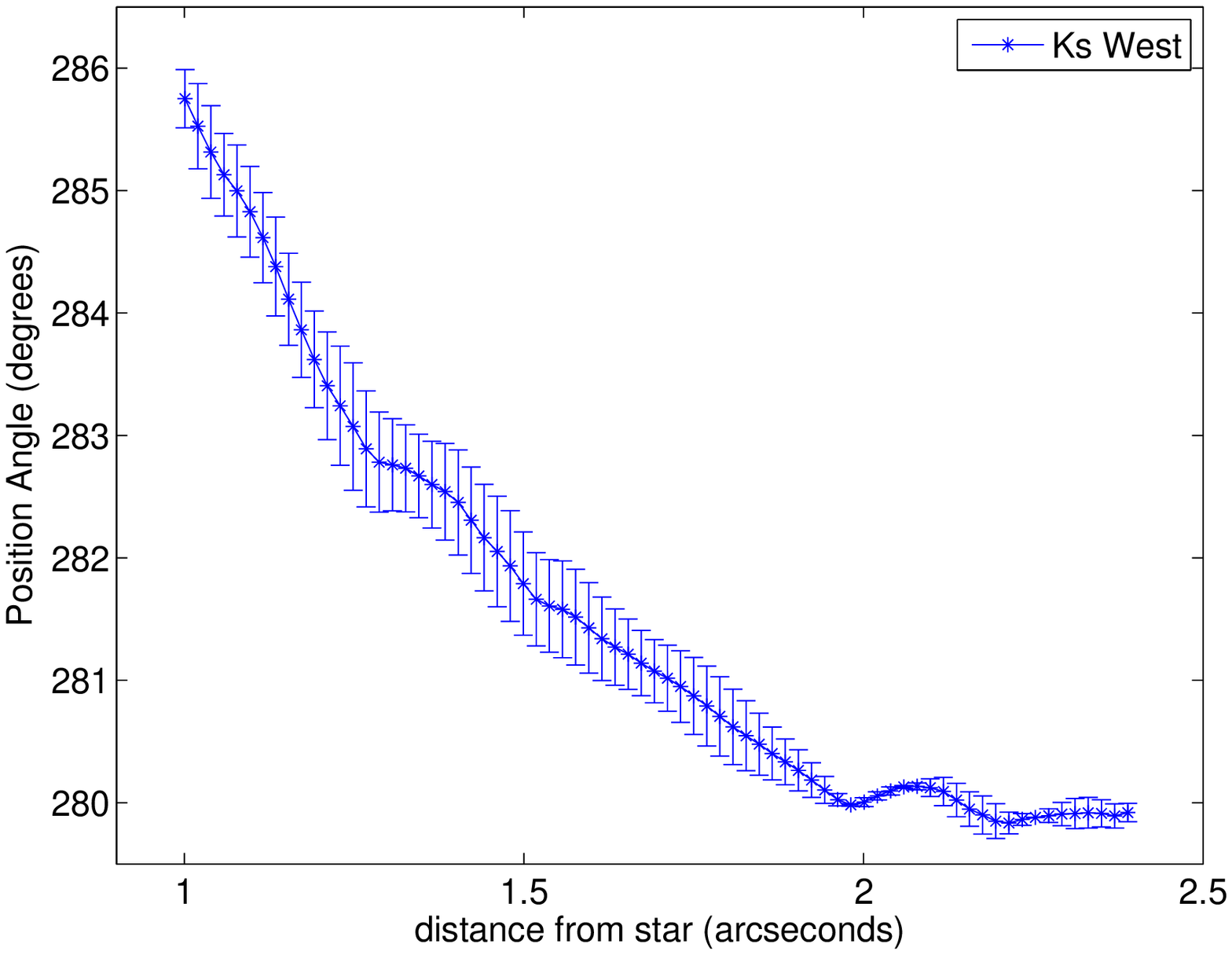}} \\
\caption{\textit{Top}: PA of the eastern side of the disk at \ks band as a function of distance from the star. See text for methodology. The PA increases with increasing distance from the star; over the same stellocentric distances a similar trend was observed at 1.1 \microns ~\citep{debes}. \textit{Bottom}: Same as above, except for the western side. The PA increases closer to the star, as is seen in the \cite{debes} 1.1 \microns ~band data.}
\end{figure}

We also measured the FWHM of the disk at \lprime . We did this in the same manner as for the \ks band image, except the \lprime ~image was binned by a factor of 4 to increase the disk SNR per pixel, and we used 3 pixel by 11 pixel (\about 0\fasec 13 by 0\fasec 47) boxes. We also added in the constant FWHM offset (\about 0\fasec 11) determined from insertion and recovery of an artificial disk in the raw data (see Appendix for details). Given the low SNRE (\about 2-3) of the \lprime ~disk and the lack of spatial coverage (detected between only 1-1\fasec 5), we do not plot the FWHM as a function of distance from the star. Instead we computed the global FWHM value as the median of the measured FWHM disk values, which was found to be 0\fasec 26 $\pm$ 0\fasec 03. The intrinsic disk FWHM, calculated by subtracting off in quadrature the PSF broadening at \lprime ~(= 0\fasec 0940), is 0\fasec 24 $\pm$ 0\fasec 03. Within the uncertainty this value is consistent with the both the intrinsic disk FWHM at \ks band, as well as the disk FWHM at 1.1 \microns ~\citep{debes} and at 0.6 \microns ~\citep{kfg}. Given that the measured FWHM is \about 2 times the size of the PSF FWHM at \lprime , the disk is marginally spatially resolved.

\subsection{Disk Position Angle}
To determine the rotational offset needed to obtain North-up, East-left in our final PISCES \ks band image, we compared our \ks band disk image to the Keck/NIRC2 \ks band image, since the rotational offset has been well-calibrated for Keck/NIRC2. We do not compare our PISCES image to published HST images to avoid any potential wavelength-dependent changes in PA. Comparing our final PISCES \ks band image with the Keck image between 1-2\asec ~revealed a difference in overall PA of 3.9\degrees , so we rotated our PISCES image by an additional 3.9\degrees ~clockwise. Fig. \ref{fig:ksimage} shows this correctly-rotated image. 

To test for any warps in the disk, we calculated the PA of the disk at \ks band on both sides of the disk as a function of distance from the star, using the disk midplane pixel locations from the Gaussian fits. The errors were measured as the difference between these values and the center of light along each 1D row at each discrete pixel distance from the star. The errors also included the 1 pixel = 0\fasec 194 astrometric centroiding uncertainty. Fig. \ref{fig:EPA} and Fig. \ref{fig:WPA} show the \ks band PA of the eastern and western sides of the disk as a function of distance from the star, respectively. The eastern PA increases with increasing distance from the star, similar to what is seen over the same spatial locations at 1.1 \microns ~\citep{debes}. The western PA increases closer to the star. This also agrees with the western PA vs. stellocentric distance seen at 1.1 \microns ~\citep{debes} and with the overall PA (278.5\degrees) measured for the western side of the disk at 0.6 \microns ~between 1.5-12\asec ~\citep{kfg}.

We calculated the PA of the disk at \lprime ~in the same manner as for the \ks band image, except we used 3 pixel by 11 pixel (\about 0\fasec 13 by 0\fasec 47) boxes. The \lprime ~image was again binned by a factor of 4. We do not plot PA vs. distance from the star due to the poor spatial coverage of the disk and, even after binning, the low SNRE (\about 2-3); however the measured values for both sides of the disk were globally consistent with the values measured at \ks band (this is also evident by examination of the \ks band and \lprime ~disk images in Fig. \ref{fig:ksimage} and Fig. \ref{fig:Limage}, respectively).

\subsection{Surface Brightness Profiles}
\begin{figure}[h]
\centering
\subfloat[]{\label{fig:KSB}\includegraphics[scale=0.45]{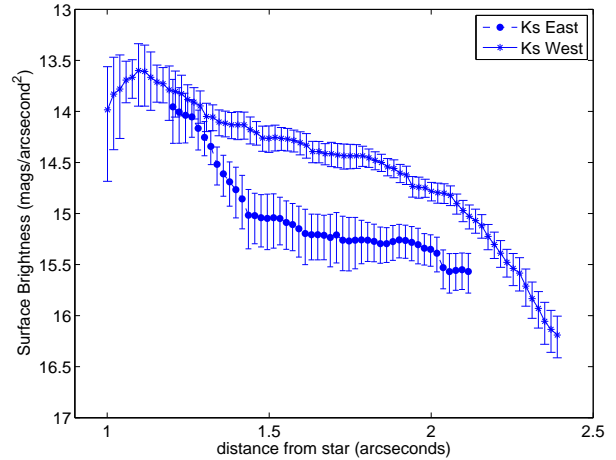}} \\
\subfloat[]{\label{fig:LSB}\includegraphics[scale=0.45]{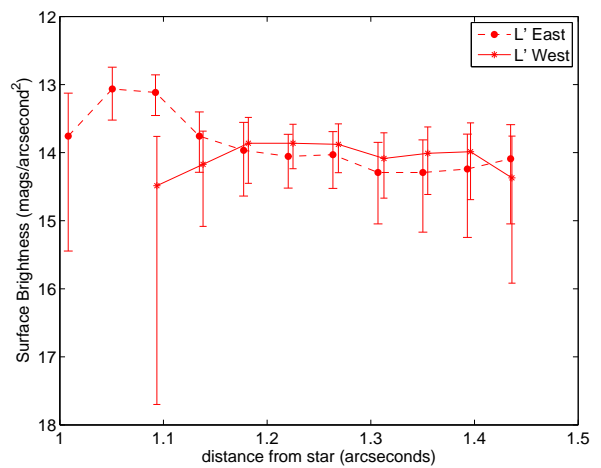}} \\
\caption{\textit{Top}: SB profile of the disk at \ks band. The western side of the disk is \about a magnitude/arcsecond$^{2}$ brighter than the eastern side at \ks band over the same stellocentric distances. The western SB profile shows a \about 2$\sigma$ drop off in SB near 1-1\fasec 2. This cannot simply be explained by flux loss due to disk self-subtraction, since this has been corrected for. \textit{Bottom}: SB profile of the disk at \lprime . Both sides of the disk are \about equally bright beyond 1\fasec 2, and just as in the \ks band image there is low SNR evidence for a decrease/flattening in SB near 1-1\fasec 2.}
\end{figure}

For both the \ks band and the \lprime ~disk images, we obtained the SB profiles by calculating the median disk flux in circular apertures with a radius of 0\fasec 15 centered on the Gaussian-fitted disk midplane location at each discrete pixel distance from the star. We chose this aperture size because it maximized the SNR of the disk flux at \lprime . This was necessary because the \lprime ~disk is detected at low SNRE (\about 2.5), so using larger apertures incorporated more background noise. We measured the errors as the standard deviation of the median SB values in the same apertures placed in a circular annulus at each pixel distance from the star. We computed the aperture correction at \ks band to be 0.64 by convolving the image of the photometric reference HBC 388 with a bar of unity counts/pixel and \about equal width to the observed disk at \ks band. The SB as a function of distance from the star is shown in Fig. \ref{fig:KSB}. ``Magnitudes" in this paper refers to Vega magnitudes.

Between \about 1.2-2\fasec 1 (where both sides of the disk are detected), the typical western disk SB is \about a magnitude/arcsecond$^{2}$ brighter than the eastern side. A similar brightness asymmetry was seen at 1.1 \microns ~\citep{debes}. It is also interesting that the western SB appears to drop off near 1\asec , by \about half a magnitude/arcsecond$^{2}$ (although the noise is high interior to this distance). We saw this SB reduction in our final ADI image as well as in our final LOCI image (Fig. \ref{fig:ksimage}), even after correction with insertion and recovery of an artificial disk. The drop in SB near 1\asec ~was not seen at shorter wavelengths \citep{kfg,debes}, which raises suspicion that the feature may not be real. However the shorter wavelength results appear to be limited by PSF residuals interior to 1\asec , so it may not be appropriate to compare to these data. 

\begin{table}[h]
\centering
\caption{\ks band SB power-law indices}
\label{tab:powers}
\begin{tabular}{c c}
\hline
Side & index  \\
\hline
West (1-1\fasec 2) & $3.48 \pm 0.5$ \\
West (1\fasec 2 $< r <$ 1\fasec 8) & $-1.75 \pm 0.05$ \\
West ($r > $~1\fasec 8) & $-4.40 \pm 0.24$ \\
East (1\fasec 2 $< r <$ 1\fasec 4) & $-5.27 \pm 0.41$ \\
East ($r > $~1\fasec 4) & $-1.1 \pm 0.07$  \\
\hline 
\end{tabular}
\end{table}

Table \ref{tab:powers} shows the power-law indices measured for the eastern and western sides of the disk at \ks band. The power-law indices do not agree very well with the indices at 1.1 \microns ~reported by \cite{debes}. The discrepancies could be explained by the differing spatial coverage of the disk at the two wavelengths. On the other hand, it is also possible that the disk SB falls differently at \ks band than at 1.1 \microns . Additional high-contrast imaging data at these wavelengths would help clarify this matter. We do not calculate the power-law indices at \lprime ~due to the lower SNRE and poor spatial coverage of the disk.

Using the \ks band power-laws, we can quantify the significance of the SB reduction interior to 1\fasec 1. From the power-law index for 1\fasec 2 $< r < $ 1\fasec 8, the expected disk SB should be 13.48 mags/arcsecond$^{2}$ at \about 1\asec. The actual value is \about  13.82, with a 1$\sigma$ upper limit of 13.64. This suggests that we are seeing a \about 2$\sigma$ reduction in SB interior to 1\fasec 1. 

We calculated the SB vs. distance from the star for the \lprime ~image in the same manner as for the \ks band image, using the same size aperture as we did for the \ks band image. Additionally the image was binned by a factor 4 to increase the signal-to-noise per pixel. We computed the aperture correction at \lprime ~to be 0.63 by convolving the photometric image of HD 15115 with a bar of unity counts/pixel and \about equal width to the measured disk at \lprime . We computed the errors in the same manner as for the \ks band SB profile. Fig. \ref{fig:LSB} shows the SB profile of the disk at \lprime . Both sides of the disk are \about equal in SB beyond 1\fasec 2, which is not seen at \ks band. Interestingly, within the uncertainties the data are consistent with a flattening in SB interior to \about 1\fasec 1, similar to the SB drop off interior to 1\fasec 1 seen in the \ks band image. 

\subsection{Limits on Planets}
\begin{figure}[h]
\centering
\subfloat[]{\label{fig:LOCI}\includegraphics[scale=0.45]{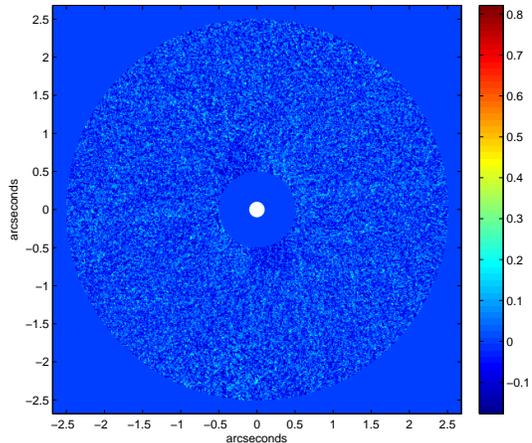}} \\
\subfloat[]{\label{fig:sensitivity}\includegraphics[scale=0.45]{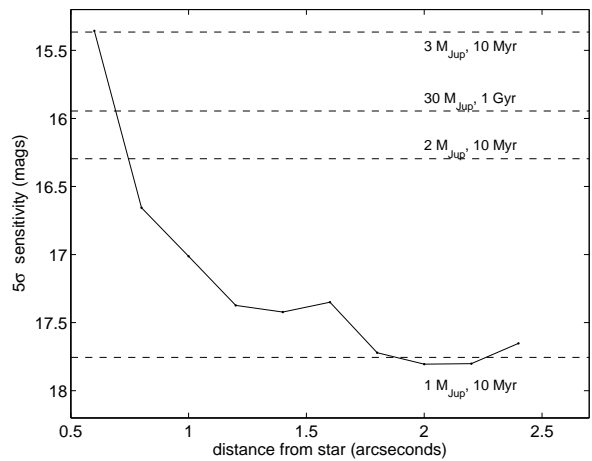}}
\caption{\textit{Top}: Final image at \lprime , in units of $\sigma$, computed with an aggressive LOCI reduction. See text for methodology. The white dot marks the location of the star and the size of a resolution element at \lprime . No high SNR point-sources are detected. \textit{Bottom}: 5$\sigma$ sensitivity curve for our \lprime ~data, computed by inserting artificial planets into the raw data and running our aggressive LOCI reduction. We reach a background limit of \about 17.8 mags, and a contrast of \about 9.5 mags at \about 0\fasec 6. The horizontal dashed lines represent the magnitudes of 10 Myr old and 10 Gyr old planets, from the \cite{baraffe} COND mass-luminosity models. We rule out planets more massive than \about 3 \mj ~if the star is young and 30 \mj ~if the star is old.}
\end{figure}

We do not detect any high SNR point-sources in any of our data reductions that would point to a possible planet. To set firm constraints on what we could have detected, we reduced the same data that was used to detect the disk at \lprime ~with an aggressive LOCI algorithm. We examined only the \lprime ~data, as opposed to including the \ks band data as well, because we had nearly double the integration time at \lprime ~(\about 40 minutes vs. 20 minutes at \ks band) and double the parallactic angle rotation (\about 40\degrees ~ vs. \about 20\degrees ~at \ks band). For the aggressive LOCI reduction we required only 0.75 FWHM of parallactic angle rotation between images, and our optimization section size ($N_{A}$) was 100. Our final image (in units of $\sigma$, where $\sigma$ was calculated in concentric annuli the size of the PSF FWHM (\about 0.\asec 1)), shown in Fig. \ref{fig:LOCI}, reveals no 5$\sigma$ point-source detections.

To ascertain the minimum flux density point-source object we could have detected, we inserted artificial planets into the raw data and re-ran our aggressive LOCI reduction. The artificial planets were bright PSFs obtained from the unsaturated images of HD 15115. We did not scale down the PSFs in brightness. Instead we simply calculated the SNR for each artificial planet's location (inside a radius = FWHM aperture) and determined the flux that would have resulted in a 5$\sigma$ detection. These are plotted in Fig. \ref{fig:sensitivity}. We reach a 5$\sigma$ background limit of \about 17.8 mags (in 40 mins of integration). At 0\fasec 6, the contrast (computed by subtracting the star's \lprime ~magnitude of 5.763) is \about 9.5 mags. These sensitivity and contrast limits show the incredible potential of direct imaging with LBTI/LMIRcam. For comparison, with Clio-2 at the 6.5 m MMT, a 5$\sigma$ background limit of 17.8 mags at \lprime ~was achieved in 2.5 hours of integration \citep{14her}.

We also overplot the magnitudes of 1-3 \mj ~planets (10 Myr old) and, conservatively, the magnitude of a 30 \mj ~brown dwarf (1 Gyr old) from the \cite{baraffe} COND mass-luminosity models (horizontal dashed lines). Assuming a young stellar age, at 5$\sigma$ confidence we rule out planets more massive than \about 1-2 \mj  ~outside of 1\asec , and outside of 0\fasec 6 we can rule out planets more massive than \about 3 \mj . If HD 15115's age is much older, then we can only rule out brown dwarfs more massive than \about 30 \mj beyond 0\fasec 7.

\section{Interpretations}
\subsection{Disk Structure}
It has been suggested that HD 15115 is interacting with the local interstellar medium (ISM), given its space motion to the south-east nearly parallel with its disk major axis PA \citep{debes32297} and the disk asymmetries seen at shorter wavelengths \citep{kfg,debes}. The \about 1 magnitude/arcsecond$^{2}$ brightness asymmetry (between 1.2-2\fasec 1) and the east-west morphological asymmetry seen in our \ks band data both support this proposition. These effects could be caused by the eastern side of the disk plowing head-first into the ISM. In this case the eastern side would be much more affected than the western side, and small grains could be blown out to the west. This could cause the eastern side to be more truncated and fainter relative to the western side, both of which we observe. Any large ($> $ a few \microns) grains in bound orbits should remain mostly unaffected. Since the disk is symmetric at \lprime , which is sensitive to large, grey-scattering grains, our data support the ISM-interaction interpretation.

It is also certainly possible that other dynamical effects (e.g., planets or a close passage of a nearby star as suggested by \cite{kfg}) are responsible for the observed morphological and SB asymmetries. However there is currently no supporting evidence for a stellar flyby \citep{kfg,debes,debes32297}. Dynamical modeling of the disk with embedded planets is beyond the scope of this paper, but such modeling would help clarify what exactly is causing the observed asymmetries in the disk.

\begin{figure}[t]
\centering
\subfloat[]{\label{fig:model}\includegraphics[scale=0.45]{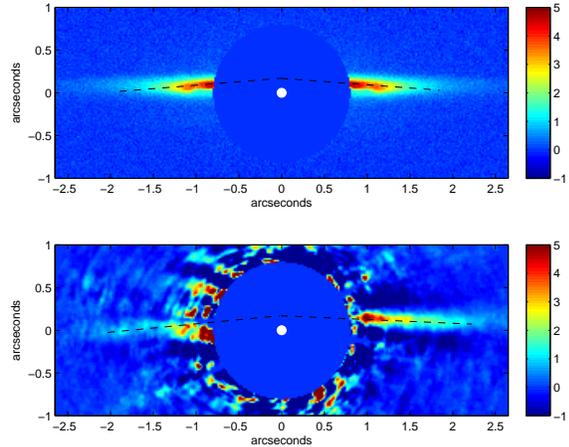}} \\
\subfloat[]{\label{fig:modelpa}\includegraphics[scale=0.45]{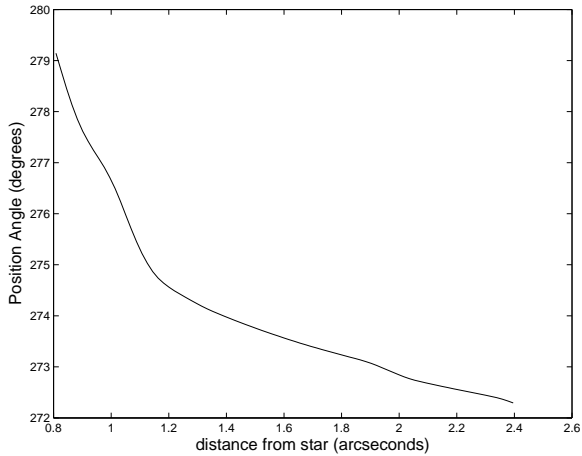}} 
\caption{\textit{Top}: Rotated model and real \ks band disk images. The model image has an inclination of 87\degrees , scattering parameter $g = 0.5$, 3 \microns ~grains, a SB power-law index of -3/2, and a gap from 0-1\fasec 1. The model image has been convolved with the PSF from the photometric image of HBC 388, and Gaussian noise has been added. Units are in \mjyasec ~for both images. The white dot marks the location of the star in both images. The dashed lines are meant to guide the eye to the apparent northern offset of the disk relative to the star. \textit{Bottom}: PA vs. stellocentric distance for the western side of the disk in the model image. The PA increases closer to the star due to the apparent bow-shape, similar to what is seen for the real data shown in Fig. \ref{fig:WPA}.}
\end{figure}

Inspection of our \ks band and \lprime ~disk images (Fig. \ref{fig:ksimage} and \ref{fig:Limage}), as well as the PA vs. stellocentric distance plots for both sides of the disk (Fig. \ref{fig:EPA} and \ref{fig:WPA}) suggests that the debris disk has a bow-like shape and is offset by a few AU to the north from the star. Specifically the offsets are \about 0\fasec 012 (0.5 AU) and 0\fasec 04 (2 AU) at \lprime , for the eastern and western sides, respectively; and \about 0\fasec 11 (5 AU) and 0\fasec 17 (8 AU) at \ks band for the eastern and western sides, respectively. Since the astrometric uncertainty in centroiding is \about 0\fasec 002 and 0\fasec 0194 at \lprime and \ks band, respectively, it is unlikely that the northern offsets could be due to centroid error. Bow-like disk morphologies have been seen in systems moving in near-perpendicular directions to their disk major axis positions (e.g., HD 61005 \citep{hd61005hines} and HD 32297 \citep{debes32297}). Because HD 15115 is moving to the south-east, nearly parallel with its disk major axis, one would not expect the disk to have a bow-shape, especially with an offset from the star to the north. However, in the HD 61005 debris disk system, \cite{hd61005maness} showed that dust grain interactions with the ISM in directions along the disk midplane could actually create bow-shapes perpendicular to the disk major axis, roughly reproducing that disk's observed morphology. Similar dynamical interactions in the HD 15115 system could be creating the observed bow-shape, though no explicit model has tested this hypothesis for this system.

Another perhaps simpler explanation for the bow-shape and the apparent offset is that these are caused by geometrical effects. We can reproduce the observed bow-shape and vertical offset with a simple inclined, ringed disk model. We used the radiative transfer equations describing the intensity of light scattering off dust particles and made three assumptions about the disk: it is not exactly edge-on, but rather inclined to 87\degrees ; it has a gap from 0-1\fasec 1~(see Section 4.3 for additional discussion of the possible disk gap); and we set the Henyey-Greenstein scattering parameter $g = 0.5$ for predominantly forward-scattering grains. This is a reasonable assumption for large ($\gtrsim $ 1 \microns ) grains scattering NIR light \citep{ggtau}. We set the dust grain size uniformly to 3 \microns ~and set the disk SB power-law index to -3/2. Fig. \ref{fig:model} shows the rotated model image, along with the rotated final \ks band image. Dashed lines have been inserted to guide the eye to the apparent bow-shape and northern offset from the star. Fig. \ref{fig:modelpa} shows the PA vs. stellocentric distance for the western side of the disk in the model image, computed in the same manner as for the real data. The PA increases closer to the star, as is observed in the real data (Fig. \ref{fig:WPA}). 

While the parameters used in our model are not a unique explanation for the observed features of the disk, the proposition that the effects can be explained by the disk's geometric orientation in space is attractive because it explains the observations without contradicting the evidence supporting the ISM interaction interpretation \citep{debes32297}. 

\subsection{Disk Color and Grain Sizes}
To determine the disk color as a function of distance from the star, we calculated $\Delta$(\ks - \lprime) = (\ks - \lprime)$_{Disk}$ - (\ks - \lprime)$_{Star}$ in the regions of spatial overlap between the \ks band and \lprime ~images (1.1-1\fasec 45). The errors were calculated by summing the individual \ks and \lprime ~errors in quadrature. The data suggest that the eastern side of the disk becomes redder than the western side with increasing distance from the star. Interior to 1\fasec 3 both sides of the disk are grey. 
\begin{figure}[h]
\centering
\includegraphics[scale=0.45]{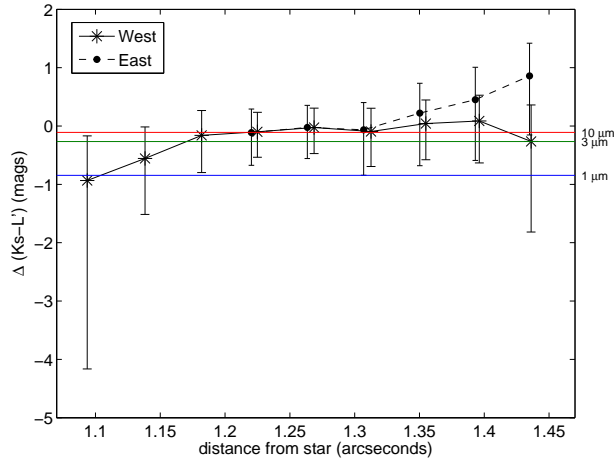}
\caption{Disk color vs. distance from the star, expressed as $\Delta$(\ks - \lprime) = (\ks - \lprime)$_{Disk}$ - (\ks - \lprime)$_{Star}$. This is calculated only where the disk is detected at both \ks band and at \lprime . To constrain dust grain size, we also plot model colors from \cite{inoue} (colored horizontal lines). The data suggest that the eastern side of the disk becomes redder than the western side with increasing distance from the star. 1-10 \microns ~grains are the best fit to the western side, while 3-10 \microns ~grains are the best match to eastern side. This may suggest that the west side of the disk is composed of smaller grains than the east side.}
\label{fig:KLcolor}
\end{figure}

To constrain the characteristic dust grain sizes, we compared our $\Delta$(\ks - \lprime) disk colors to the models of \cite{inoue}. These models calculate disk colors in the NIR and assume silicate dust grain composition. The modeled grain sizes range from 0.1 \microns ~to 10 \microns ~and assume a single grain size for the dust population. We plot the model disk colors for 1, 3, and 10 \microns ~grains. These are shown as the colored lines in Fig. \ref{fig:KLcolor}. We also modeled 0.1 \microns ~and 0.3 \microns ~grains, but their colors were too blue to be supported by the data.

From Fig. \ref{fig:KLcolor}, we see that 1-10 \microns -sized grains are the best match to the western side of the disk, while 3-10 \microns ~are the best match to the eastern side. This suggests that the observed scattered light from the disk is tracing the large parent body dust grains in the disk. The data also may suggest that the west side of the disk is composed of smaller grains than the east side. The blowout grain size for this system is expected to be \about 1-3 \microns ~\citep{hahnblowout}, assuming a stellar mass of 1.3 $M_{\odot}$, a luminosity of 3.3 $L_{\odot}$, and a grain density in the range 1-2.5 g/cm$^{3}$. Grains smaller than the blowout size would be blown out radially; on the eastern side, these grains would hit the ISM and be blown back to the west, resulting in the western side being dominated by smaller blue-scattering grains. Our observational constraints on the dust grain sizes offer some support for these predictions.

\subsection{Does the disk have a gap?}
The SB profiles at \ks band and at \lprime ~drop off or flatten out (at low SNR) near 1\asec . For an edge-on disk with no gap, the SB should continue to increase closer to the star. Because we do not see this in our data, this may be an indication that the disk has a gap interior to \about 1\fasec 1. This is consistent with prediction of a gap near 1\asec ~by \cite{moordisks2011}, using spectral energy distribution (SED) analysis of the system. Though degenerate with temperature and dust grain size, their best-fit model of HD 15115's infrared SED yields a two-component disk, with the inner at 4 AU and the outer at 42 AU (both $\pm 2$ AU). Given a distance to the star of 45.2 pc, the outer edge of the gap would be at 0\fasec 93 $\pm$ 0\fasec 04. This agrees well with the observed drop offs in SB near 1\asec . 
\begin{figure}[h]
\centering
\includegraphics[scale=0.45]{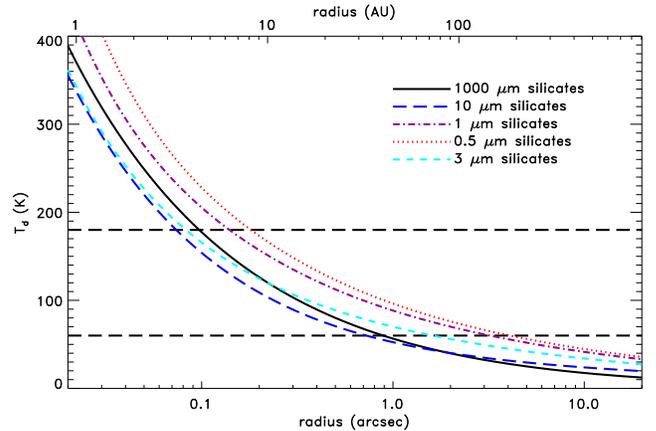}
\caption{Equilibrium disk temperature vs. distance from the star, for several different silicate grain sizes. See text for methodology. The horizontal dashed lines represent temperature constraints of 179 K and 57 K from \cite{moordisks2011}. Given that we have observational evidence for the predicted gap at \about 1\asec , we can independently constrain the dust grain sizes to be \about 3 \microns ~for astronomical silicates.}
\label{fig:tvsr}
\end{figure}

We can independently constrain the parent body dust grain size using estimates of the disk's equilibrium temperature as a function of distance from the star, for different grain sizes (Fig. \ref{fig:tvsr}). The equilibrium dust temperature was computed by balancing the absorption and the emission energy of a particle with adopted dust properties for astronomical silicates \citep{laor93}. Under an optically thin condition, the heating is solely from the central star, representative as the best-fit Kurucz model with a stellar temperature of 7000 K and a stellar luminosity of 3.3 L$_{\odot}$. 

The horizontal dashed lines represent the predicted disk temperatures of 179 K and 57 K from \cite{moordisks2011}. Assuming we have observationally detected the inner edge of the outer disk component (and hence the gap) at \about 1\fasec 1, the parent body dust grain sizes are constrained to be \about 3 \microns . This is consistent with our dust grain size estimates from the disk colors.

\subsection{Limits on a planet inside the gap}
\begin{figure}[h]
\centering
\includegraphics[scale=0.45]{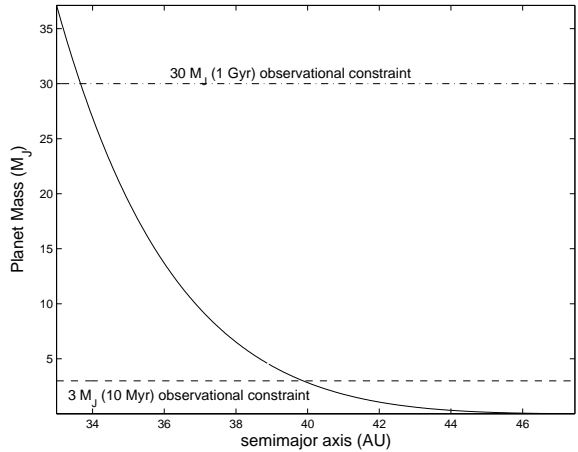}
\caption{Masses of a possible object orbiting inside a disk gap with its outer edge at 1\fasec 1 (50 AU), as a function of semimajor axis, computed using Eq. \ref{eq:chaotic} (solid line). The dashed line represents our 3 \mj ~(10 Myr old age) observational constraint and assumes that the planet's semimajor axis = its projected separation at the epoch of our observations. The dash-dot line is the same, except showing our 30 \mj ~(1 Gyr old age) observational constraint. If HD 15115 is old, then the object creating the gap must reside between \about 34-45 AU. If the system is young, the allowed semimajor axis range shrinks to 40-45 AU.}
\label{fig:chaotic}
\end{figure}

We can also calculate an independent estimate of the mass of a planet creating the gap in the disk using the equation describing the relationship between the width of the chaotic zone around an assumed coplanar, low-eccentricity planet and its mass and semimajor axis \citep{malhotra98}: 
\begin{equation} \label{eq:chaotic}
\Delta a \approx 1.4~a_{p}~(M_{p}/M_{*})^{2/7},
\end{equation} 
where $\Delta a$ is the width of the chaotic zone, $a_{p}$ and $M_{p}$ are the semimajor axis and mass of the planet, respectively, and $M_{*}$ is the mass of the star. We make the assumption that, from our SB profiles, the outer edge of the gap, $r_{gap}$, is at 1\fasec 1 (50 AU). Since $\Delta a = r_{gap} - a_{p}$, we can substitute this into Eq. \ref{eq:chaotic} and solve for the planet mass as a function of semimajor axis (Fig. \ref{fig:chaotic}). 

The planet's mass approaches zero the closer it orbits to the disk edge and approaches the brown dwarf mass regime closer to the star. Assuming the gap was created by an object whose semimajor axis = its projected separation at the epoch of our observations and that the system is young, we can say that the object must be $\lesssim$ a few \mj ~because we could have detected $\gtrsim$ 3 \mj ~planets (dashed line in Fig. \ref{fig:chaotic}). The planet's semimajor axis would then be constrained to be between 0.87-1\asec ~(40-45 AU). If the system is old, the object must be less than \about 30 \mj ~and its semimajor axis would be constrained to 0.75-1\asec ~(34-45 AU).

It is certainly possible that we did not detect the putative planet because it is currently in front of or behind the star. It is also possible that the gap has not been created by a planet, but instead arises from other dynamical interactions. Additional high-contrast, high angular resolution imaging of the debris disk would help constrain the existence of both the gap and any possible planets.

\section{Summary}
We have presented several intriguing results on the debris disk surrounding HD 15115. These results are:
\begin{enumerate}
  \item We detect east-west asymmetry in the disk morphology at \ks band, with the western side of the disk being a \about magnitude/arcsecond$^{2}$ brighter than the eastern side at the same stellocentric distances. The asymmetry and brightness differences are consistent with results at shorter wavelengths and lend additional evidence to the interpretation that ISM interactions are affecting the disk structure \citep{debes32297}.
  \item At \lprime , we detect symmetric and \about equally bright disk structure. 
  \item We detect an overall bow-like shape to the disk at both \ks band and \lprime , and the disk appears offset from the star to the north by a few AU. We are able to explain these observed effects using a model disk with a near edge-on inclination, a gap between 0-1\fasec 1, large (3 \microns ) grains, and mostly forward-scattering grains ($g = 0.5$).
  \item The FWHM of the disk at \ks band and at \lprime ~is consistent with the disk FWHM at 1.1 \microns ~\citep{debes} and at 0.6 \microns ~\citep{kfg}.
  \item The disk SB profile at \ks band shows evidence for a 2$\sigma$ reduction interior to 1\fasec 1 (50 AU). Combined with SED analysis, this may be a sign that the disk has a gap interior to 1\asec ~(45 AU). Additional high-contrast observations at NIR wavelengths with better inner working angles would help confirm or disprove the proposition that the disk has a gap.
  \item The \ks - \lprime ~disk color is mostly grey for both sides of the disk between 1.1-1\fasec 45 (50-66 AU). 1-10 \microns ~grain sizes are the best match to the western disk color. 3-10 \microns ~grain sizes are the best match for the eastern side. Given the system's expected grain blowout size of 1-3 \microns , our dust grain size constraints may support the ISM interaction interpretation \citep{debes32297}, which predicts that small grains would be blown to the western side of the disk, leaving large, unaffected grains on the eastern side. SED analysis, combined with our observational evidence for a gap near 1\asec ~(45 AU), also predicts a dust grain size of \about 3 \microns . 
  \item We do not detect any $\gtrsim$ 5$\sigma$ point-sources at \lprime ~indicative of planets. We rule out companions more massive than \about 3 \mj ~beyond 0\fasec 6 if the star is young and more massive than 30 \mj ~beyond 0\fasec 7 if the star is old. Independently we constrain the mass of a single, coplanar, low-eccentricity planet creating the gap in the disk to be \about a few \mj ~if it is close to the disk edge, and to be in the brown dwarf regime if its orbit is closer to the star. Assuming the object's semimajor axis = its projected separation at the epoch of our observations, the planet would be $\lesssim$ 3 \mj ~and orbit between 0.87-1\asec ~(40-45 AU) if the star is young; if the system is old, the object would be less massive than \about 30 \mj ~and orbit between 0.75-1\asec ~(34-45 AU). 
\end{enumerate}

\acknowledgments
We thank Piero Salinari for his insight, leadership and persistence which made the development of the LBT adaptive secondaries possible. We thank the LBTO staff and telescope operators for their hard work facilitating use of the telescope and instruments. We are grateful to Elliott Solheid, the lead mechanical engineer on the PISCES project. Roland Sarlot and Andrew Rakich provided support in the PISCES optical design and engineering, respectively. We acknowledge support for LMIRcam from the National Science Foundation under grant NSF AST-0705296. We thank A.K. Inoue and M. Honda for sharing their disk modeling data and for helpful discussions. We thank John Debes, the referee, for helpful discussions and for his careful review of the manuscript. T.J.R. acknowledges support from the NASA Earth and Space Science Graduate Fellowship.

\clearpage
\appendix
\section{Accounting for Disk Self-Subtraction by LOCI}
\begin{figure}[t]
\centering
\subfloat[]{\label{fig:ksexpected}\includegraphics[scale=0.43]{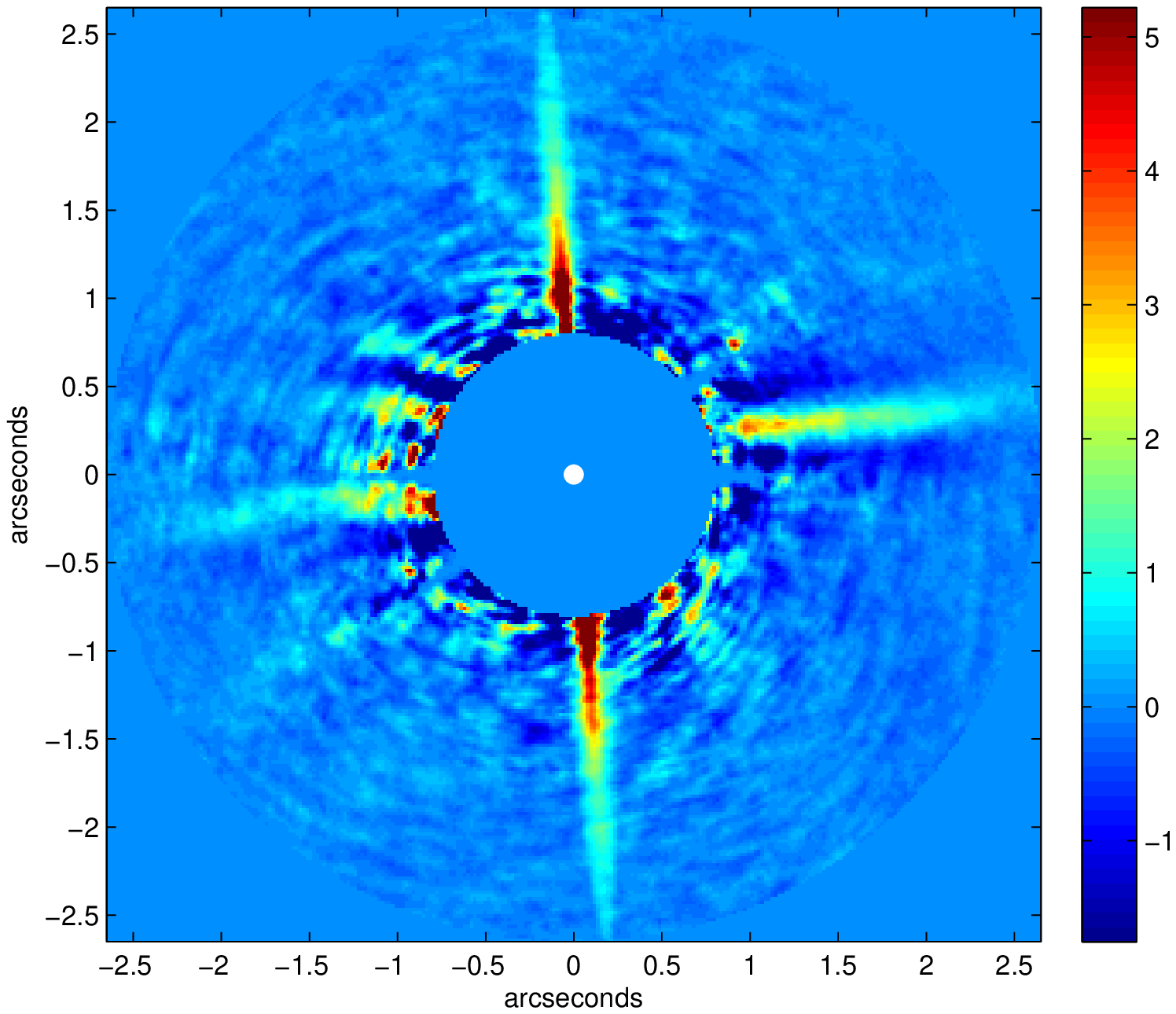}} 
\subfloat[]{\label{fig:ksobserved}\includegraphics[scale=0.43]{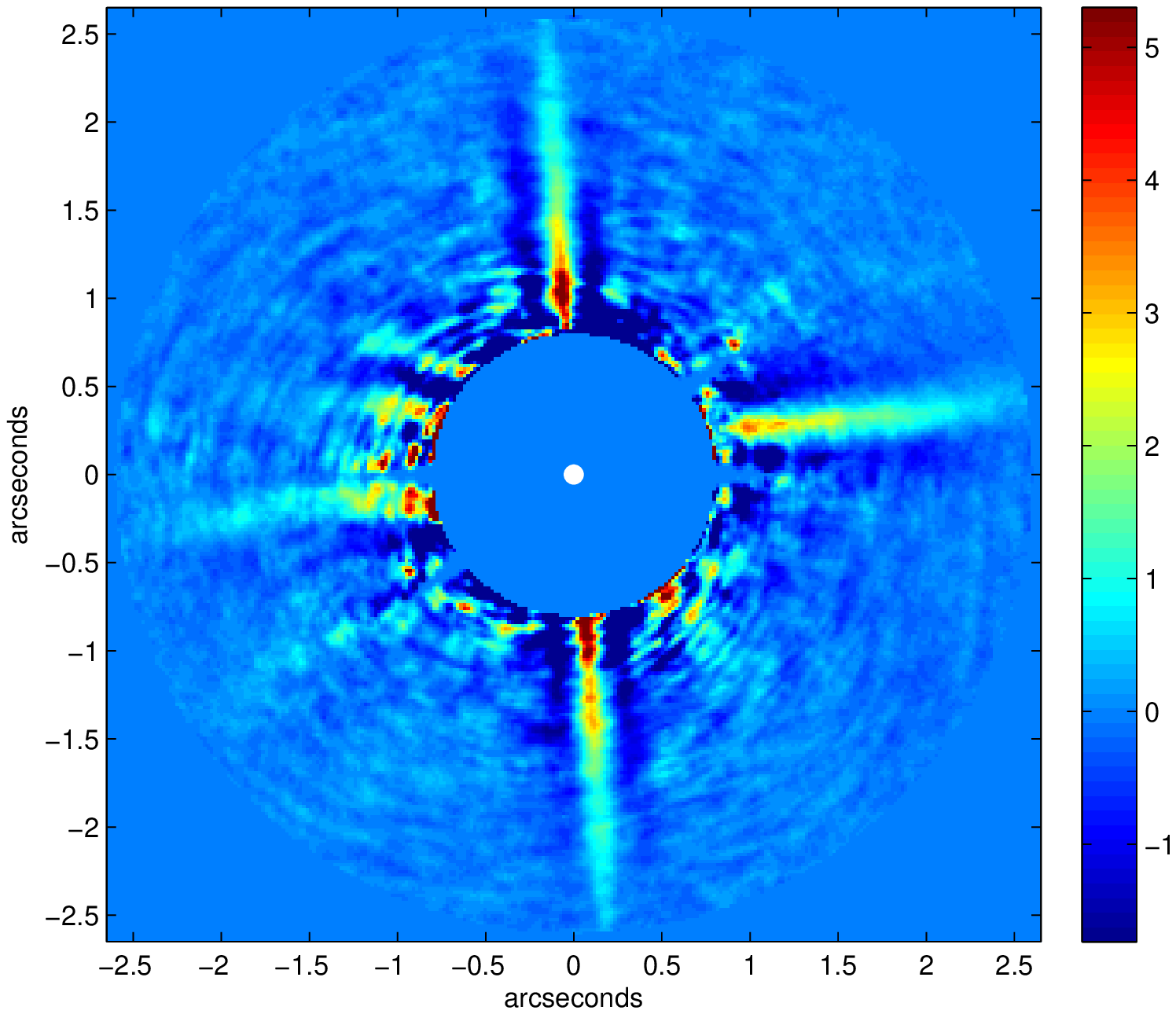}} \\
\subfloat[]{\label{fig:Lexpected}\includegraphics[scale=0.43]{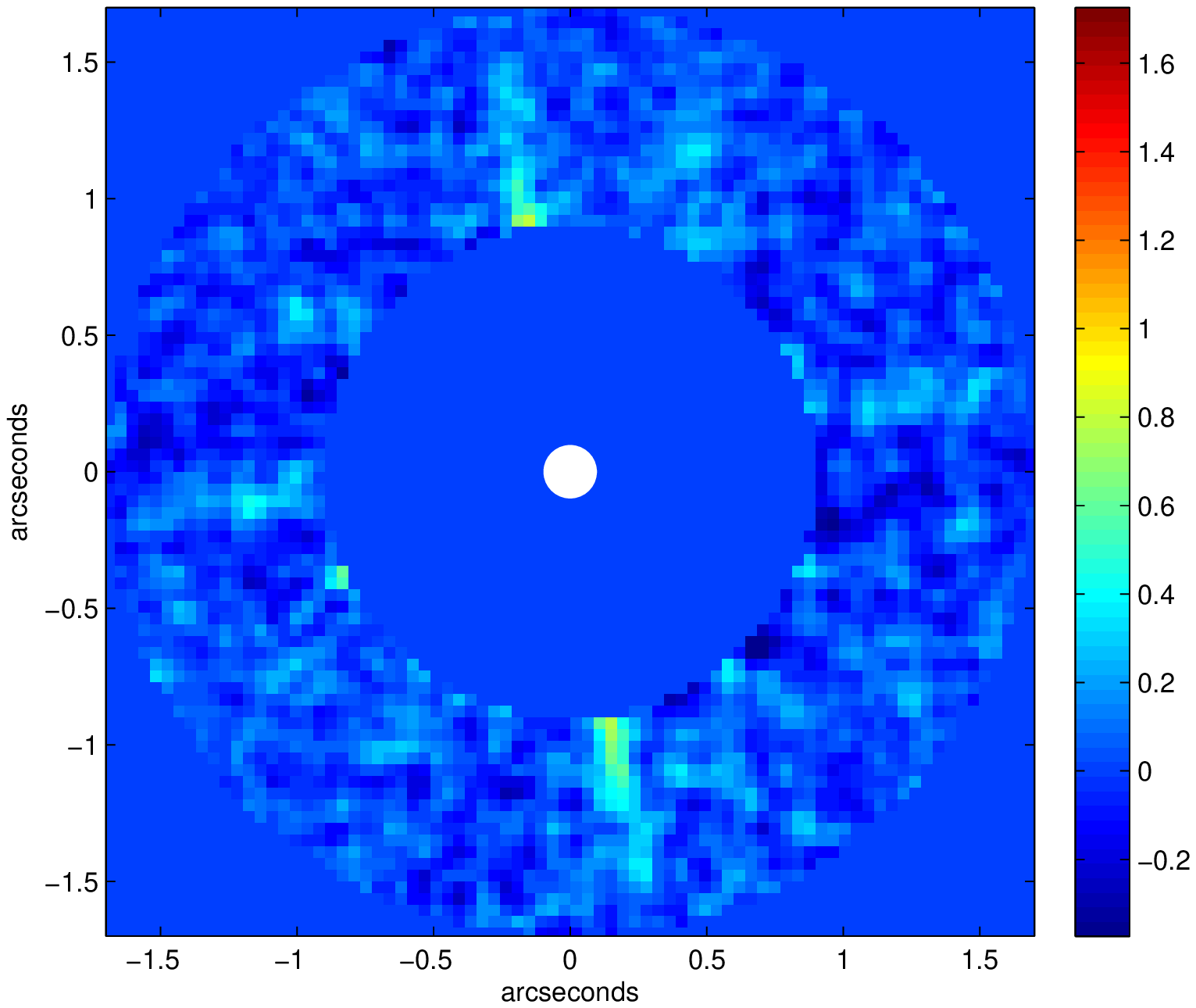}} 
\subfloat[]{\label{fig:Lobserved}\includegraphics[scale=0.43]{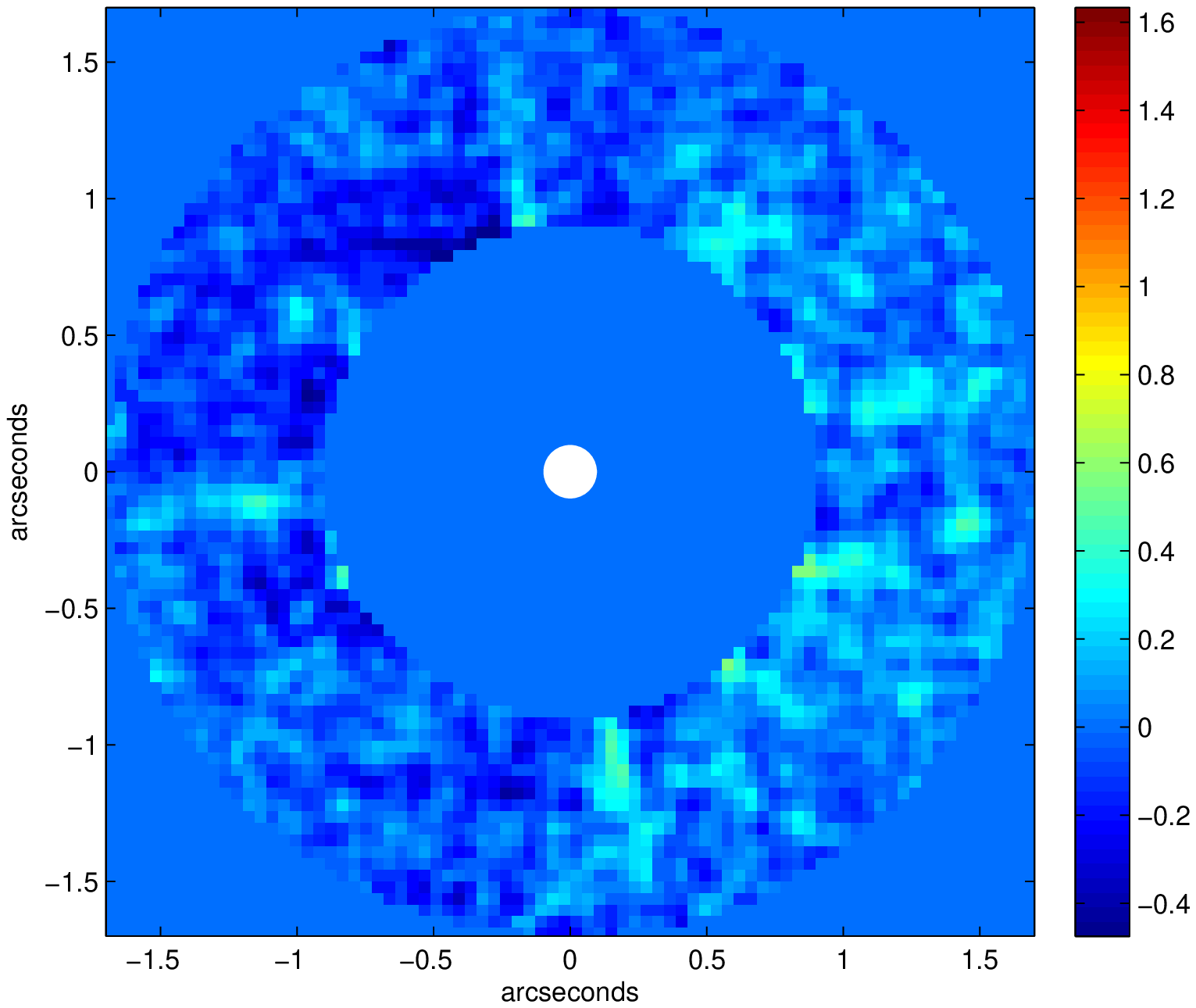}}
\caption{Expected and observed artificial disks at \ks band and \lprime . Model disks are always vertical. \textit{Top left}: \ks band image of the artificial disk, showing what the disk should look like if unaltered by the LOCI algorithm reduction. \textit{Top right}: \ks band image of the recovered model disk. \textit{Bottom left}: the same as (\textit{a}) except at \lprime . \textit{Bottom right}: the same as (\textit{b}), except at \lprime . In all cases, the model disks are recovered, though some self-subtraction is evident.}
\end{figure}

To account for disk self-subtraction by the LOCI algorithm, at each wavelength we inserted an artificial model disk into the raw data, at a PA nearly perpendicular to the known disk, ran the data through the pipeline, and recovered the model disks. The artificial disks were set to SB levels comparable to the real disks at each wavelength, and the widths of the artificial and real disks were comparable. Fig. \ref{fig:ksexpected} and Fig. \ref{fig:ksobserved} show the expected and recovered model disks at \ks band, respectively; Fig. \ref{fig:Lexpected} and Fig. \ref{fig:Lobserved} show the same at \lprime . 

\begin{figure}[t]
\centering
\subfloat[]{\label{fig:Kspacorrect}\includegraphics[scale=0.43]{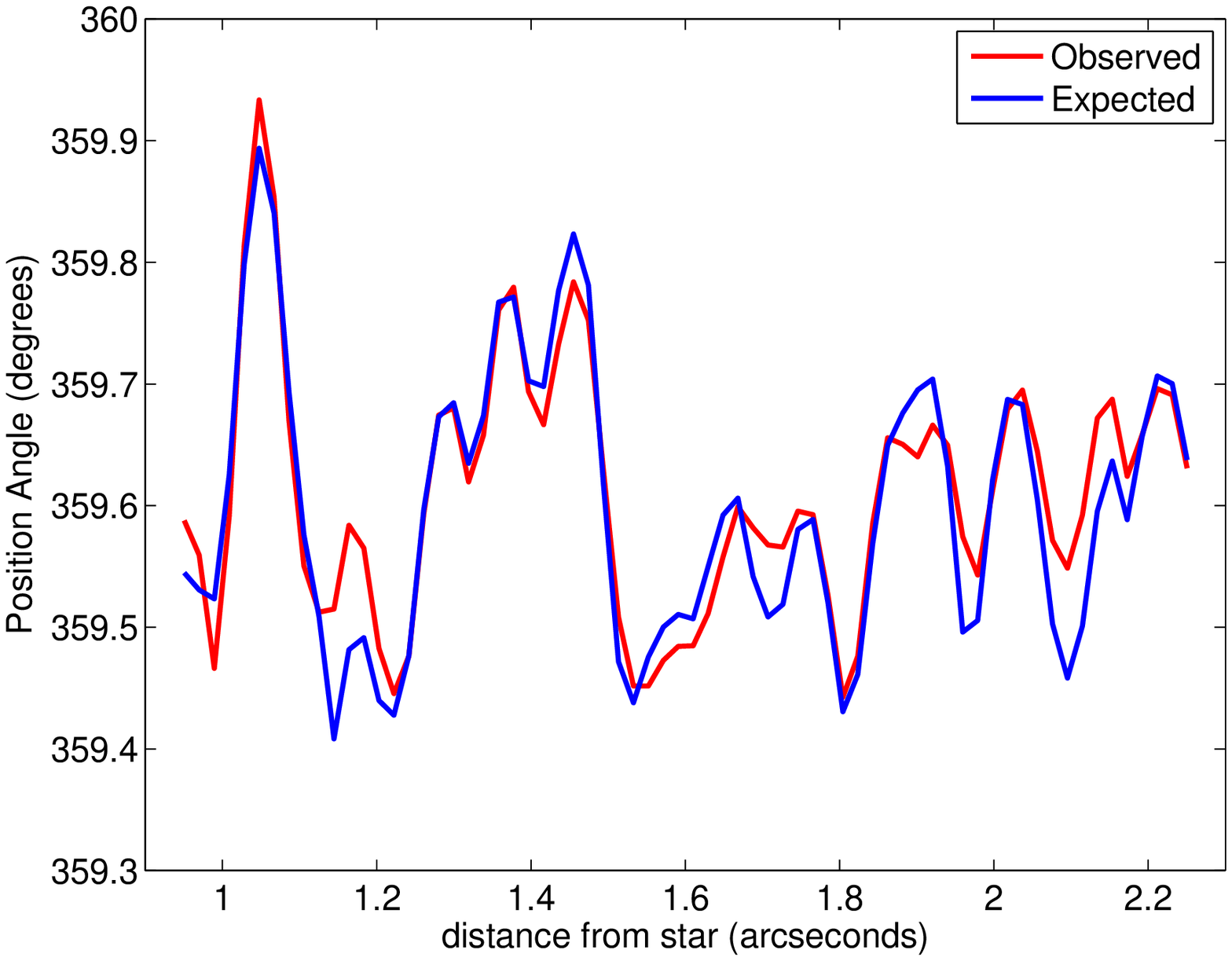}} 
\subfloat[]{\label{fig:Lpacorrect}\includegraphics[scale=0.43]{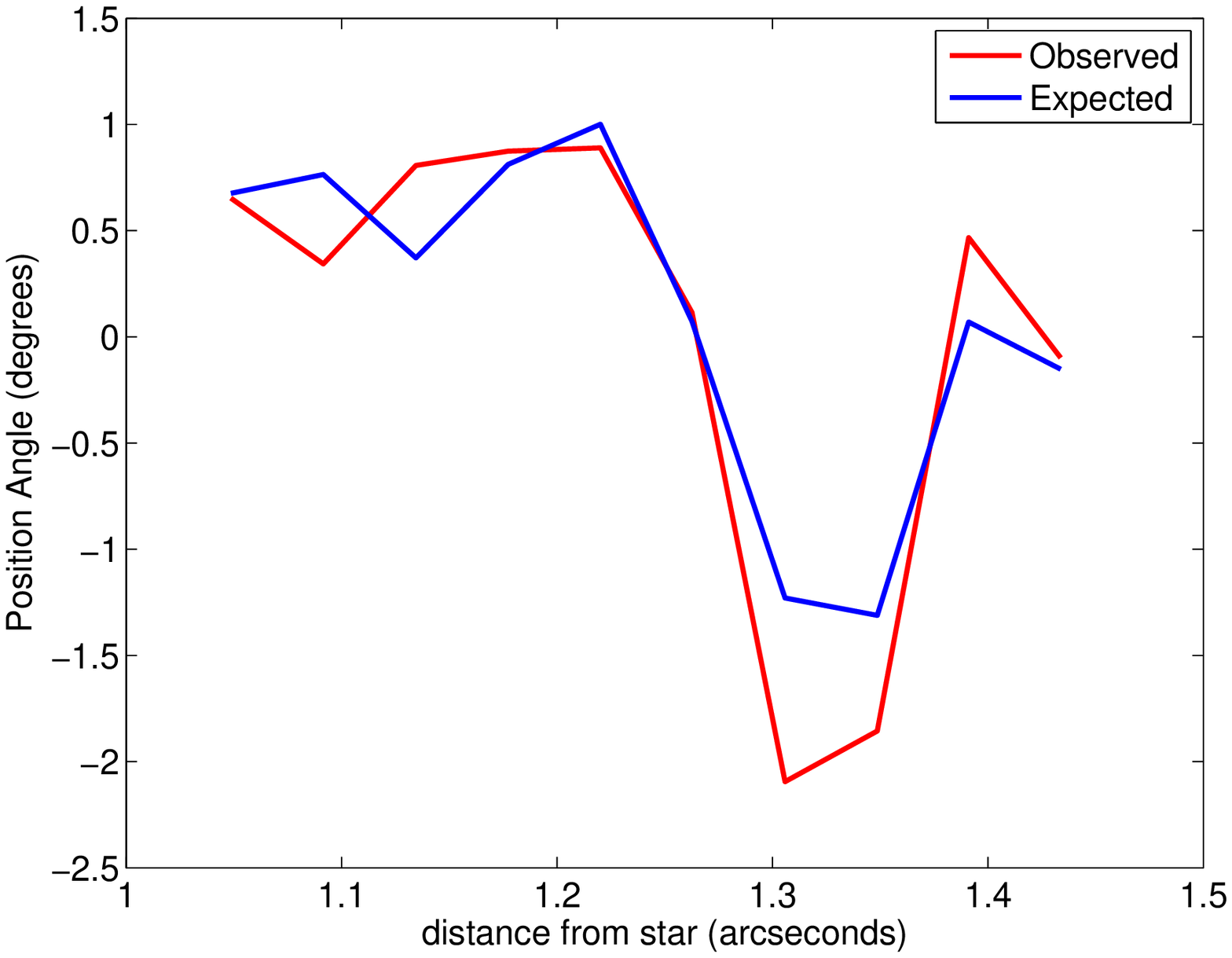}} \\
\subfloat[]{\label{fig:Ksfwhmcorrect}\includegraphics[scale=0.43]{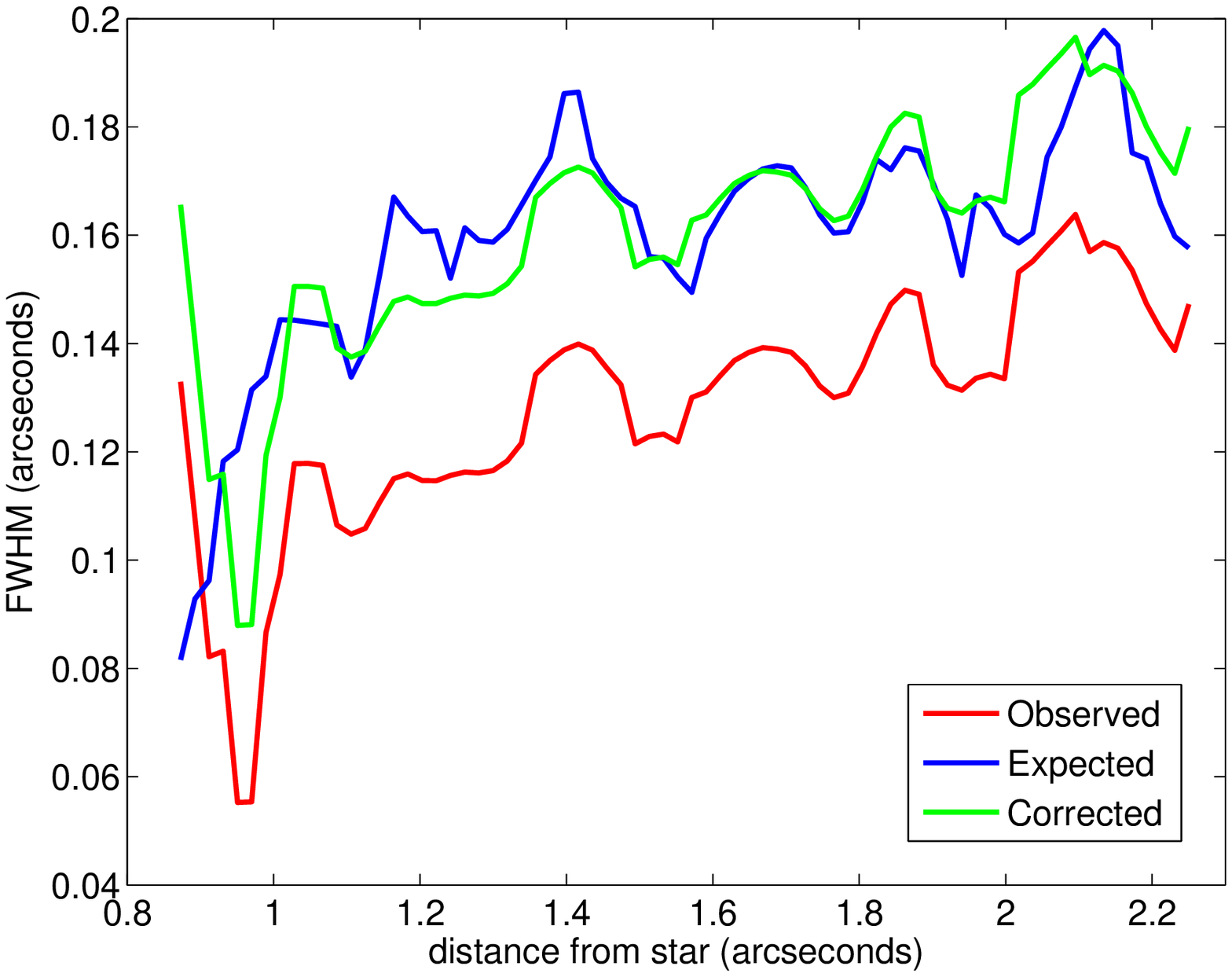}} 
\subfloat[]{\label{fig:Lfwhmcorrect}\includegraphics[scale=0.43]{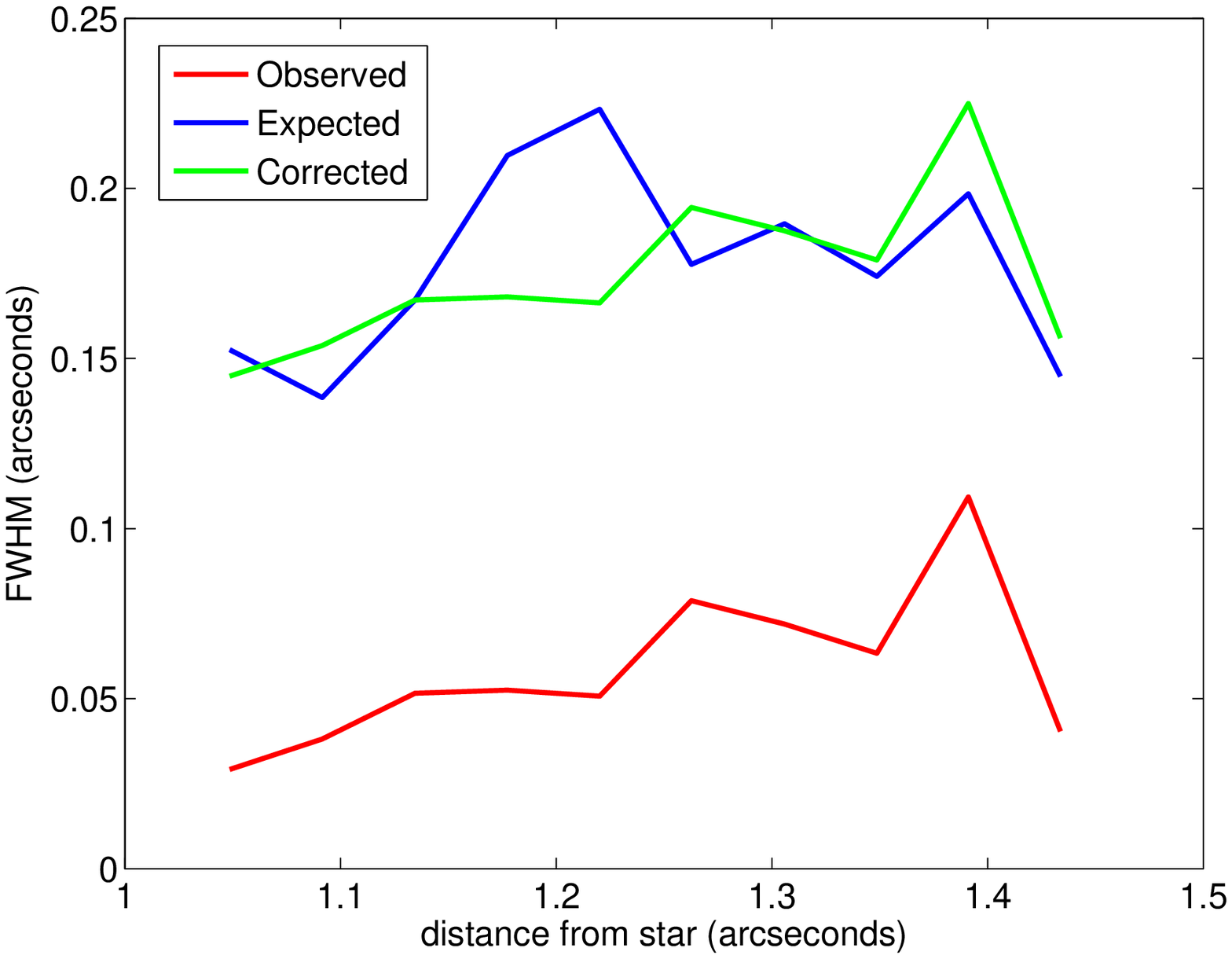}} \\
\subfloat[]{\label{fig:Kssbcorrect}\includegraphics[scale=0.43]{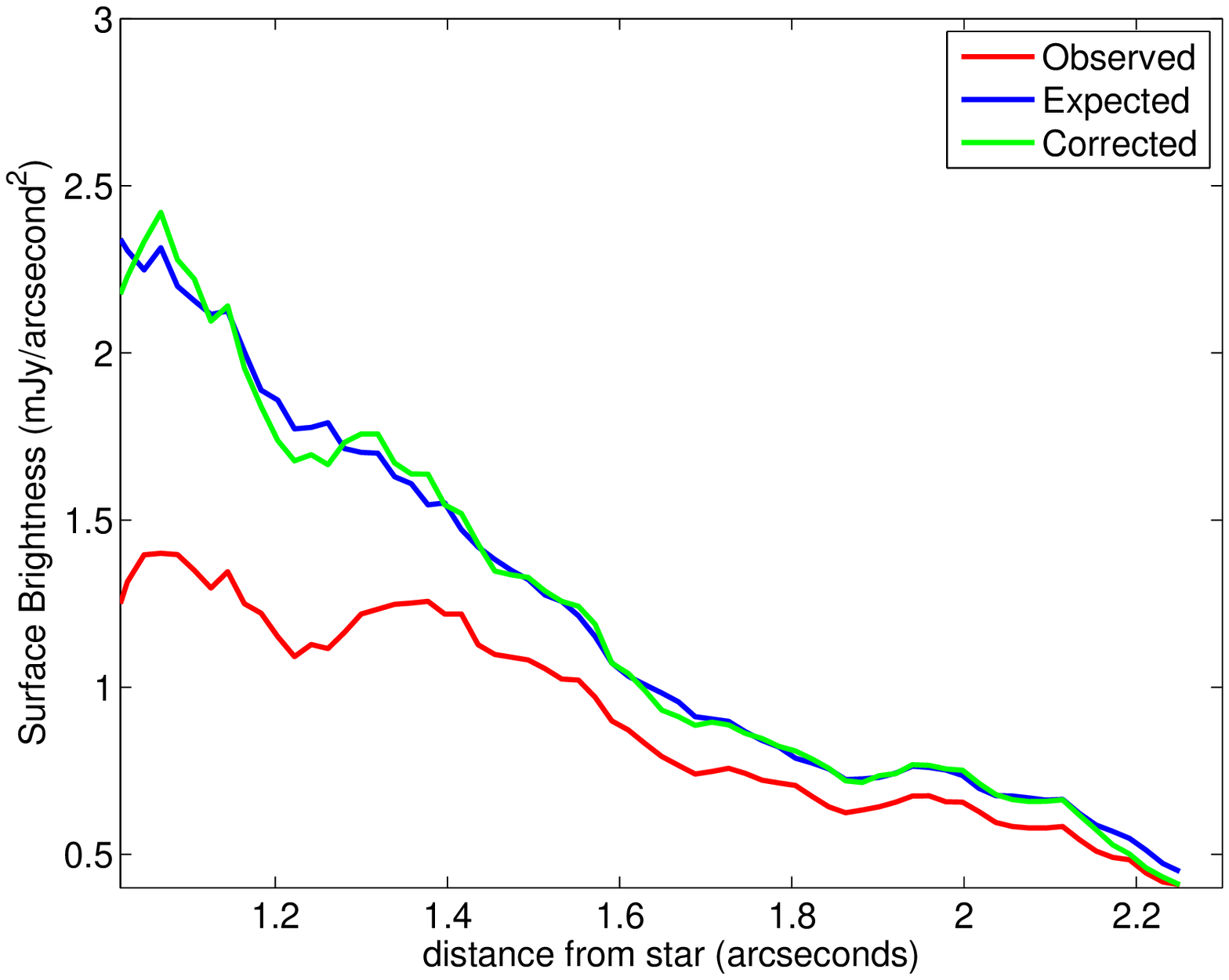}} 
\subfloat[]{\label{fig:Lsbcorrect}\includegraphics[scale=0.43]{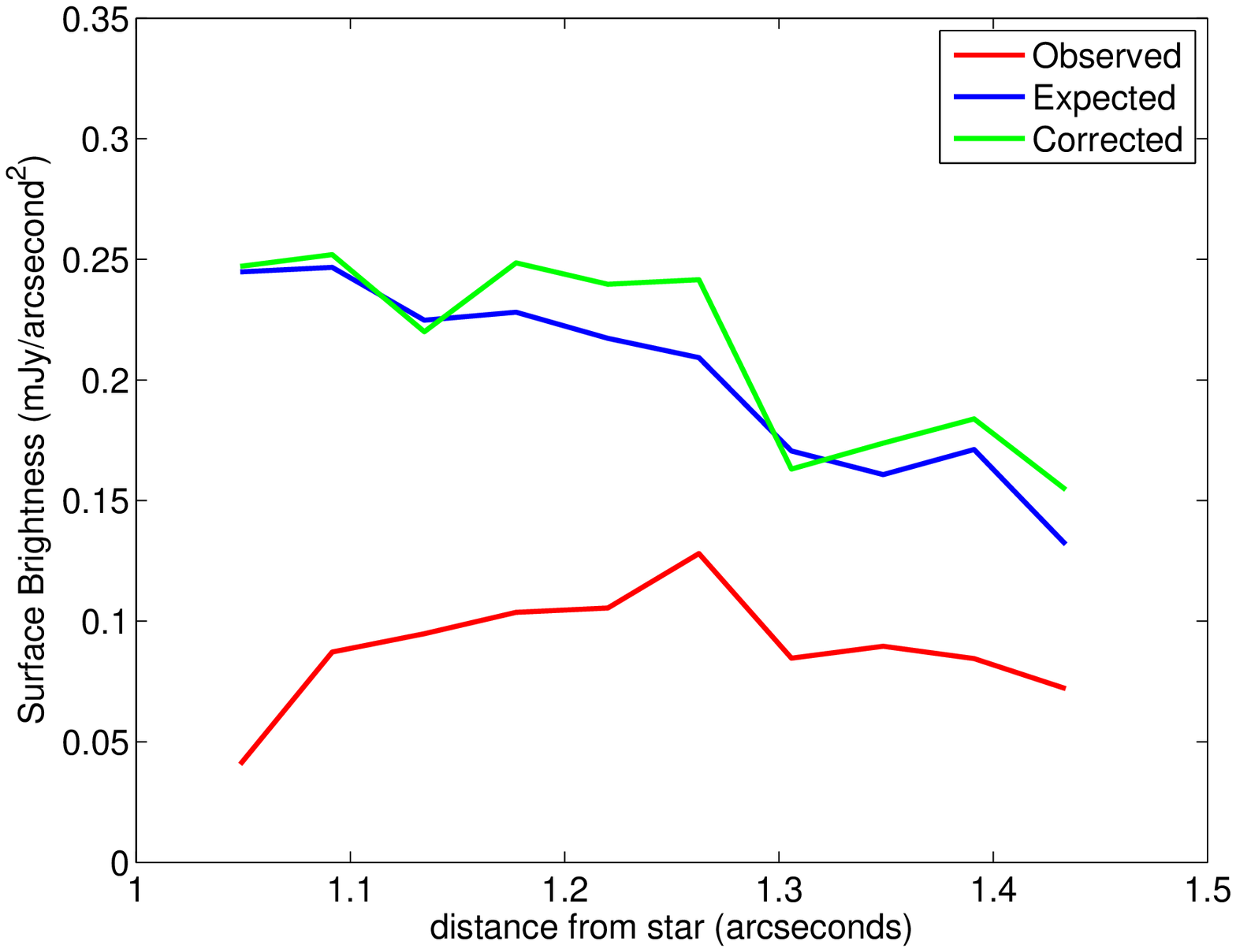}} 
\caption{\textit{Top left}: Expected (blue) and observed (red) model disk PA vs. distance from the star at \ks band. No correction is needed. \textit{Top right}: the same, except at \lprime . No correction is needed. \textit{Middel left}: Expected (blue), observed (red), and corrected (green) model disk FWHM vs. distance from the star at \ks band. An offset correction of \about 0\fasec 03 is needed. \textit{Middle right}: the same, except at \lprime . An offset correction of \about 0\fasec 11 is needed. \textit{Bottom left}: Expected (blue), observed (red), and corrected (green) model disk SB vs. distance from the star. \textit{Bottom right}: the same, except at \lprime .}
\end{figure}

After recovering the model disks, we compared the observed PA, FWHM, and SB values as a function of distance from the star with the expected PA, FWHM, and SB values. The calculations were performed with identical methods to the real disk data analysis. At both wavelengths, the expected and observed PA values are consistent with each other, therefore no correction was needed (see Fig. \ref{fig:Kspacorrect} and Fig. \ref{fig:Lpacorrect}). At \ks band, an addition to the observed FWHM of \about 0\fasec 03 was needed to correct the apparent constant offset (see Fig. \ref{fig:Ksfwhmcorrect}). At \lprime , a correction was also needed, with a value of \about 0\fasec 11 (see Fig. \ref{fig:Lfwhmcorrect}). Both FWHM correction offsets have been included in the FWHM analysis of the real disk images.

To correct for the reduction in disk SB at both wavelengths relative to the expected values, we mapped out the self-subtraction as a function of distance and multiplied this into each recovered disk. The corrected SB values, along with the expected and observed, are shown in Fig. \ref{fig:Kssbcorrect} (\ks band) and Fig. \ref{fig:Lsbcorrect} (\lprime ). The SB corrections have been included in the SB analysis of the real disk images.

\clearpage

\bibliographystyle{apj}
\bibliography{ms}


\end{document}